\documentclass[10pt,sigconf,letterpaper,nonacm]{acmart}

\usepackage{amsmath}
\usepackage{booktabs}
\usepackage{enumitem}
\usepackage{graphicx}
\usepackage{multirow}
\usepackage{soul}
\usepackage{subcaption}
\usepackage{tabularx}
\usepackage[most]{tcolorbox}
\usepackage{tikz}
\usepackage[table]{xcolor}

\setcopyright{acmlicensed}
\copyrightyear{2026}
\acmYear{2026}
\acmDOI{XXXXXXX.XXXXXXX}
\acmConference[Conference acronym 'XX]{Make sure to enter the correct
  conference title from your rights confirmation email}{June 03--05,
  2018}{Woodstock, NY}
\acmISBN{978-1-4503-XXXX-X/2018/06}

\author{Jiaming Tang}
\affiliation{%
  \institution{University of Michigan}
  \city{Ann Arbor}
  \state{MI}
  \country{USA}
}
\email{jmtang@umich.edu}

\author{Chenlan Wang}
\affiliation{%
  \institution{University of North Florida}
  \city{Jacksonville}
  \state{FL}
  \country{USA}
}
\email{chenlan.w@unf.edu}

\author{Mingyan Liu}
\affiliation{%
  \institution{University of Michigan}
  \city{Ann Arbor}
  \state{MI}
  \country{USA}
}
\email{mingyan@umich.edu}

\author{Armin Sarabi}
\affiliation{%
  \institution{University of Michigan}
  \city{Ann Arbor}
  \state{MI}
  \country{USA}
}
\email{arsarabi@umich.edu}

\begin{document}

\title{Decoding the Legalese: A Scalable and Quantitative Framework for Analyzing Corporate Privacy Policies}

\begin{abstract}
Even though privacy policies are the primary mechanism organizations use to disclose how they collect, process, and share personal data, they are difficult for average users to interpret, perhaps by design, due to their verbosity and dense legal language. Importantly, there is a lack of standardized metrics that characterize key qualities of a privacy policy beyond regulatory requirements. Recent advances in large language models (LLMs) make it feasible to automatically structure and analyze these documents at scale. In this study, we develop and evaluate an end-to-end, LLM-enabled system that converts raw privacy policies into fine-grained structured representations and a set of quantitative measures. Our pipeline applies a detailed taxonomy to extract specific data elements and governing practices, capturing relational links that connect each practice to the data elements it references. We apply our framework to a diverse corpus of 10,000 website privacy policies, yielding, to the best of our knowledge, the most comprehensive dataset of its kind to date. Building on our structured representations, we introduce the first standardized and repeatable quantitative metrics for evaluating privacy policies along four dimensions: completeness, transparency, commitment to user protection, and emphasis on business-driven data practices. This allows us to compare policies within and across industry sectors, and to assess the tension between user protection and business interests.
\end{abstract}

\maketitle

\section{Introduction}
In today's data-driven digital society, individuals' personal information is routinely collected, processed, and shared by corporations. One's privacy risks are no longer limited to whether a company has in its possession one's data; instead, risks often arise when data from different contexts is combined to infer sensitive traits, when automated systems use these inferences to make decisions about individuals or groups, and when data is reused or redistributed across third-party partners in ways that users cannot easily see or control. This expanded collection and sharing also increases exposure to data breaches, where stored or transmitted information is accessed or leaked without authorization, potentially leading to harms like identity theft, fraud, or targeted scams.

Within this context, privacy policies are the primary mechanism companies use to disclose their data collection and related practices, yet they remain difficult for most people to understand and interpret. Across industries, policies use inconsistent terminology, vary in completeness, and rely on dense, often vague legal language shaped by uneven legal and compliance pressures. This, compounded by consent fatigue, makes it difficult for users to determine what data is collected about them, why it is collected, how it is processed and retained, who it is shared with, and what rights they have with respect to their data. Moreover, users, and even researchers and domain experts, often struggle to assess an organization's commitment to protecting consumer data (e.g., relative to industry norms) versus how aggressively it collects and monetizes that data for business purposes.

Rigorous analysis of privacy policies often requires converting \emph{unstructured} natural language into \emph{structured} formats (e.g., JSON), a process that has historically been labor-intensive, slow, costly, and error-prone~\cite{wilson2016creation,ravichander2019question,zimmeck2019maps,arora2022tale}.
Recent advances in LLMs make it increasingly feasible to automate privacy policy annotation and analysis at scale. 
Early studies demonstrated that LLMs can effectively replace or augment traditional rule-based and supervised natural language processing (NLP) pipelines for large-scale annotation of privacy policies, achieving competitive performance with minimal task-specific training~\cite{tang2023policygpt,rodriguez2024large,sun2024empowering}. Subsequent research has emphasized prompt engineering to elicit structured representations, enabling LLMs to extract practices and reason over policy semantics without extensive fine-tuning~\cite{huang2024analyzing,goknil2024privacy}. In parallel, 
LLMs have been applied to map privacy policy statements to modern requirements such as General Data Protection Regulation (GDPR) and California Consumer Privacy Act (CCPA) while identifying potential violations~\cite{xie2025evaluating}. More recently, LLMs have supported privacy risk interpretation and user-facing explanations~\cite{chen2025clear}. Collectively, these efforts position LLMs as increasingly powerful tools for structured privacy policy analysis and interpretation.

Despite a growing body of work on LLM-based privacy policy analysis, prior research has not closed the loop by translating qualities of a privacy policy into concise, user-facing metrics.
This study aims to close this gap by developing and evaluating an end-to-end system that takes a raw privacy policy as input and outputs first-of-their-kind quantitative metrics that jointly characterize and assess a privacy policy. Our main contributions are as follows:

\begin{itemize}[leftmargin=*,itemsep=0pt]
    \item We develop an end-to-end system that substantially extends prior approaches for producing standardized, fine-grained structured representations of privacy policies. The pipeline applies a detailed taxonomy to extract the data elements an organization collects, the practices governing their use, and the relational links between them, such as which practices apply to which data elements.
    \item We apply this framework to a diverse corpus of 10,000 website privacy policies, producing, to the best of our knowledge, the most comprehensive data of its kind to date in both scale and representational richness.
    \item Building on this foundation, we define four quantitative metrics that systematically aggregate fine-grained representations into high-level measures of a privacy policy's completeness, transparency, commitment to user protection, and the organization's propensity to monetize or otherwise derive value from users' data, resulting in a novel, repeatable scoring framework for privacy policies.
    \item We apply this scoring framework to our corpus of website privacy policies to understand population-wide characteristics as well as differences across industry sectors, and shed light on questions such as where corporate and user interests are most in tension, and what might constitute a ``good'' privacy policy.
\end{itemize}

In the remainder of this paper, we present related work (\autoref{sec:background}), our extraction pipeline (\autoref{sec:method}) and its evaluation and validation (\autoref{sec:evaluation}), followed by the analysis framework (\autoref{sec:analysis}) and results (\autoref{sec:results}). \autoref{sec:conclusion} concludes the paper.

\section{Background and Related Work}\label{sec:background}
Before instruction-tuned LLMs made zero- and few-shot approaches practical for privacy policy analysis, most methods relied on supervised models trained on manually annotated corpora, e.g.,~\cite{wilson2016creation,zimmeck2019maps,arora2022tale}. The manual effort required to build these datasets constrained both their scale and the granularity of their taxonomies. For example, \cite{wilson2016creation} uses 15 data categories to characterize user data collected by an organization. Subsequent work leveraging LLMs expanded this taxonomy to 34 categories and 125 descriptors of individual data elements. By contrast, in this study we further extend it to 237 data descriptors, along with additional enhancements described below. \cite{alamo2022systematic,javed2024systematic,van2024privacy} provide broader reviews of prior work on privacy policy analysis, predating the widespread adoption of instruction-tuned LLMs due to their strong zero-/few-shot performance and improved reliability in producing faithful structured representations.

With the advent of modern LLMs, recent work has increasingly automated privacy policy annotation and analysis, including compliance assessment~\cite{xie2025evaluating,zhao2025llm}, topic labeling~\cite{tang2023policygpt,goknil2024privacy,silva2024entailment}, extracting mentions of personal data types~\cite{rodriguez2024large}, evaluating reliability for segment classification, question answering, and summarization~\cite{sun2024empowering}, and producing structured annotations of first-party collection and third-party sharing statements~\cite{yang2025automated,zhao2025llm}. Perhaps the closest prior work to ours is \cite{huang2024analyzing}, which develops an end-to-end system for structuring privacy policies using an extended taxonomy based partly on \cite{wilson2016creation}. More broadly, LLMs have been used to extract structured representations across security~\cite{siracusano2023time,hu2024llm,oates2024using,kravsovec2025large,sarabi2025ransomware}, systems and network measurement~\cite{sarabi2023fingerprint,jiang2024lilac,zhong2024logparser}, scientific literature analysis~\cite{dagdelen2024structured}, legal analysis~\cite{savelka2023unreasonable}, and clinical domains~\cite{huang2024critical,reichenpfader2024scoping}, underscoring the broad applicability of LLM-driven pipelines for converting free text into standardized formats.

With the exception of \cite{huang2024analyzing} and \cite{xie2025evaluating}, prior work utilizing LLMs for privacy policy analysis primarily benchmark LLMs (e.g., against previously established ground truth such as \cite{wilson2016creation}) and typically target only one or a few stages of the analysis pipeline (e.g., labeling pre-chunked segments, which corresponds to the first step of our pipeline in \autoref{sec:section-discovery}). In contrast, we develop an end-to-end system that starts from an Internet domain and produces a structured representation of the domain's privacy policy under a more comprehensive, custom-designed taxonomy. More importantly, to the best of our knowledge, our work is the first to use these structured extractions to characterize the broader landscape and to define novel quantitative metrics for evaluating both the quality of privacy policies and their orientation toward user protection versus business interests. This focus complements compliance-oriented approaches such as \cite{xie2025evaluating}, which evaluate privacy policies against requirements derived from regulations such as the GDPR~\cite{gdpr2016}. While compliance checks are essential for identifying critical omissions and deficiencies, our metrics and analysis also allows us to quantify the inherent tension between corporate interests and consumer/user protection, seen through the lens of privacy policies.  

From a technical standpoint, we extend the taxonomy of \cite{huang2024analyzing} in both breadth and granularity and, under this expanded schema, build what we believe is the largest dataset to date of structured privacy policy representations, spanning small and large organizations across multiple industry sectors. Our taxonomy captures the major technical facets of privacy policies: data collection, purposes of data collection/processing, data retention, data protection, data sharing, and user rights. Beyond the increase in coverage, we also introduce and extract explicit relational structure that (1) connects data collection statements to subjects and collection methods/sources, (2) connects data sharing statements to recipients and stated purposes, and most importantly, (3) connects each stated data practice to the specific data elements it governs. While prior work incorporates subsets of these components, our system is the first to integrate them all into a single end-to-end pipeline that outputs a holistic, structured representation of a privacy policy across the major topics it covers.

\section{A Structuring Pipeline}\label{sec:method}
In this section, we describe our end-to-end pipeline that automatically retrieves a website's privacy policy and transforms it into a standardized, machine-readable JSON representation based on a predefined schema and taxonomy.

\subsection{Pipeline Overview}

\begin{figure}[t]
  \centering
  \includegraphics[width=\linewidth]{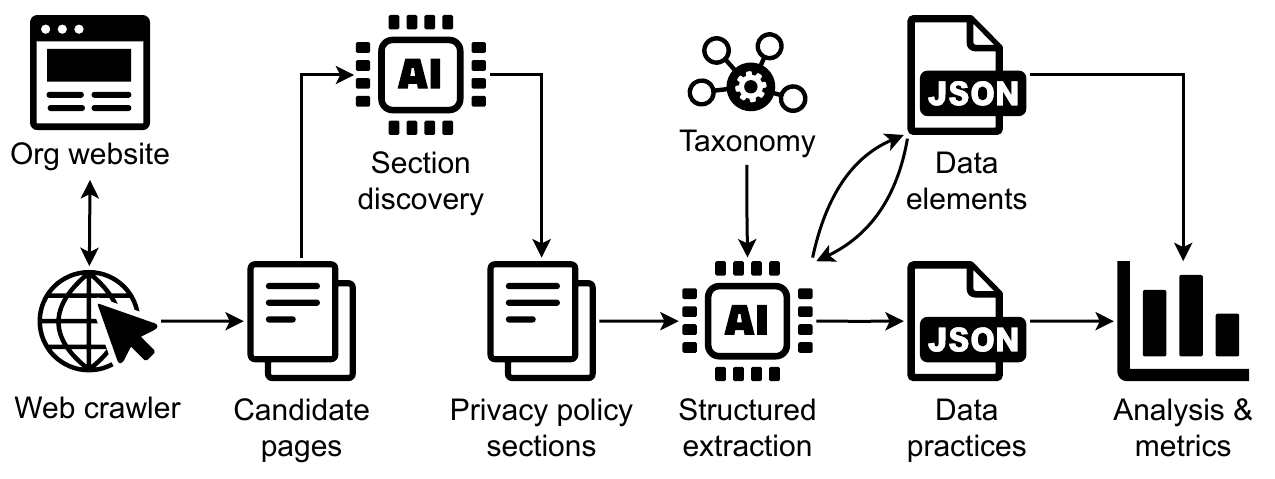}
  \caption{High-level overview of our privacy policy retrieval and structuring pipeline.}
  \label{fig:pipeline-overview}
\end{figure}

\autoref{fig:pipeline-overview} shows our end-to-end pipeline, which includes: (1) a web crawler for retrieving pages likely to contain a website's privacy policy (\autoref{sec:retrieval}); (2) extraction and segmentation of privacy policy content into topical sections (\autoref{sec:section-discovery}); (3) schema-constrained, taxonomy-driven extraction of data elements collected by the organization (\autoref{sec:data-element}); and (4) structured extraction of data practices with semantic links to data elements that they govern (\autoref{sec:data-practice}).

Our pipeline relies on two key components: (1) a detailed, expert-defined taxonomy inspired by prior work~\cite{wilson2016creation,huang2024analyzing} and extended to support the extraction of fine-grained attributes and their relationships; and (2) task-specific LLM prompts that operationalize this taxonomy to produce structure policy representations. We develop both components through an iterative refinement process: we evaluate the pipeline on representative sample text, analyze its outputs, and then revise the taxonomy to improve coverage and specificity while strengthening prompt instructions to reduce common errors. We also incorporate few-shot learning by augmenting each prompt with manually annotated text from real privacy policies, selected to capture nuanced cases. Examples of our prompt are provided in \autoref{app:prompt_specifications}.

In the remainder of this section, we describe our pipeline in detail. Although our taxonomy is inspired by \cite{huang2024analyzing} (which itself is partially based on \cite{wilson2016creation}), we make several extensions. We improve section discovery, substantially expand the controlled vocabulary for collected data elements, and add additional contextual attributes (e.g., data subjects, data collectors, and collection methods) to support more granular outputs. More importantly, we semantically link data practice statements to the specific data elements and subjects they govern, enabling the fine-grained quantitative analysis presented in \autoref{sec:analysis}. We also conduct a blind validation of LLM outputs (\autoref{sec:validation}), in contrast to the non-blind validation used in \cite{huang2024analyzing}, to better assess output fidelity and agreement with human-generated annotations.

\subsection{Policy Retrieval and Preprocessing}\label{sec:retrieval}

We develop our own privacy policy retrieval mechanism rather than relying on existing large-scale datasets~\cite{amos2021privacy,wagner2023privacy}. This allows our framework to be applied to any Internet domain and, more importantly, to capture an up-to-date snapshot of the privacy policy landscape, which is necessary because policies can change over time, as shown in \autoref{sec:tiktok}. To retrieve webpages that may contain a privacy policy, we adopt the crawling logic of \cite{huang2024analyzing}, implemented using Crawlee~\cite{crawlee} with headless Chromium. The crawler first follows up to three links from the bottom of the homepage containing the word ``privacy'' and two common paths (\texttt{/privacy-policy} and \texttt{/privacy}). From these seed pages, it then follows up to five additional ``privacy'' links from the top of each page to discover policies hosted behind privacy hub/center pages. \cite{huang2024analyzing} reports that these heuristics successfully retrieve the privacy policy of roughly 90\% of companies in the Russell 3000 index. The remaining cases are largely due to websites that do not provide a privacy policy page (or provide only a PDF) or crawler-related failures (e.g., timeouts or bot detection), rather than failures in identifying privacy policy pages. While coverage could likely be improved, e.g., by incorporating LLM-guided navigation, our goal is to obtain a representative corpus of privacy policies for our analysis, for which a small error rate is acceptable, thus crawler improvements is not a focus of the present study.

We convert HTML of successfully retrieved candidate pages (HTTP status $<400$) to Markdown using Docling~\cite{docling}, detect language using Lingua~\cite{lingua}, and discard non-English pages. Next, we remove boilerplate by filtering text spans longer than 100 characters that also appear on the website's homepage, eliminating common navigation text and cookie consent banners. We also discard pages with fewer than 5,000 characters after boilerplate removal, as manual inspection showed that these are typically trivial policies from organizations with little or no data collection (e.g., only IP addresses of website visitors). We therefore exclude policies shorter than this threshold to focus on organizations whose data collection warrants more comprehensive privacy policies.

\subsection{Section Discovery}\label{sec:section-discovery}

To detect the presence of a privacy policy and to avoid passing irrelevant text into subsequent steps, we classify and partition candidate pages as follows. \cite{huang2024analyzing} uses cues from HTML tags (e.g., headings) to identify section boundaries. However, we found that labeling a segment based only on its heading can miss other secondary topics discussed in the section body, e.g., a  ``what information do we collect'' section may also describe purposes of data collection without explicitly stating that in the heading. We therefore provide the full page content to the LLM and instruct it to (1) determine whether a privacy policy is present, and if so, segment it into logical sections; and (2) assign each section one or more primary and secondary labels. Primary labels capture the section's main topic (as indicated by the heading or dominant content), while secondary labels capture additional topics discussed in the body but not its main focus. Primary and secondary labels are selected from a set of topics, largely based on \cite{huang2024analyzing}, including \texttt{elements} (collected data elements), \texttt{methods} (data collection methods), \texttt{purposes} (purposes or legal bases for collection), \texttt{handling} (retention and security mechanisms), \texttt{sharing} (whether and how data is shared with third parties), \texttt{rights} (user rights, choices, and controls), \texttt{cookies and tracking} (cookies, web beacons, and other tracking technologies), and \texttt{audiences} (information for specific audiences, such as users in particular jurisdictions). For this study we have excluded sections labeled \texttt{audiences} from subsequent processing as they indicate audience-specific policies, which are a subject of future work.

\subsection{Extraction of Collected Data Elements}\label{sec:data-element}

\begin{table}[t]
    \caption{Data element categories grouped by meta-category. Numbers in parentheses indicate the number of explicit data descriptors in our taxonomy for each category/meta-category.}
    \label{table:data-categories}
    \footnotesize
    \centering
    \begin{tabular}{@{} p{\linewidth} @{}}
    \hline
    \textbf{Data (meta-)categories} \\
    \hline
    \textbf{Identity/background (58)}: personal identifier (11), contact info (3), demographic info (16), education info (6), professional info (13), vehicle info (9) \\
    \textbf{Digital profile (39)}: online identifier (8), account info (10), device info (15), network info (6) \\
    \textbf{Physical/health profile (26)}: physical characteristic (5), medical info (9), biometric data (6), fitness \& activity data (6) \\
    \textbf{Financial/legal profile (38)}: financial info (7), financial standing (13), insurance info (8), legal info (10) \\
    \textbf{Physical behavior (24)}: precise location (4), approximate location (4), movement data (4), sensory data (6), on-site data (6) \\
    \textbf{Digital behavior (50)}: product \& service usage (8), Internet usage (8), content generation (5), transaction data (5), communication data (4), feedback data (4), diagnostic data (4), configuration data (7), preference data (5) \\
    \textbf{Other (2)}: other (2) \\
    \hline
    \end{tabular}
\end{table}

Next, we identify data elements that the organization states it collects. Note that while an LLM can be naively instructed to annotate phrases that state specific data elements that are collected, this can lead to inconsistent extraction granularity and make it difficult to align synonymous data elements. We therefore implement data element extraction as a structured annotation task constrained by a predefined schema and an extensive taxonomy, building on the taxonomy developed by \cite{huang2024analyzing}. We concatenate all sections labeled \texttt{elements}, \texttt{methods}, or \texttt{cookies and tracking} (as either primary or secondary labels) and provide the combined text to the LLM. We then instruct the LLM to extract phrases denoting collected data elements and to infer five attributes for each extraction: \texttt{category}, \texttt{descriptor}, \texttt{subjects}, \texttt{collectors}, and \texttt{methods}. \texttt{category} maps each extraction to one of 33 categories under 7 meta-categories (\autoref{table:data-categories}), while \texttt{descriptor} provides a finer-grained specification using a controlled vocabulary of 237 descriptors. If no descriptor applies, the LLM may generate its own, allowing us to iteratively expand the vocabulary by inspecting and incorporating new data elements. The remaining fields capture the subject from whom the data is collected, the entity that initially collects it, and the collection method; valid values for these fields are listed in \autoref{table:schema} in \autoref{app:tables}.

Compared to \cite{huang2024analyzing}, our schema for extracting collected data elements captures additional contextual attributes (\texttt{subjects}, \texttt{collectors}, and \texttt{methods}). Additionally, while we largely retain the same set of high-level data categories, we substantially refine and expand the descriptor vocabulary from 125 to 237 descriptors, resulting in 13\% out-of-vocabulary extractions in the final dataset.

\subsection{Extraction of Data Practices}\label{sec:data-practice}

In addition to collected data elements, we extract and structure statements about data practices, including data collection purposes, data handling (retention policies and protection mechanisms), data sharing, and user rights (choices available to users and access rights concerning their data). These practices can be specified in relation to particular data rather than stated globally. For example, indefinite retention periods often apply only to anonymized or aggregated data. Organizations may also adopt different practices for sensitive versus non-sensitive data. Disambiguating these cases by semantically linking practices to their related data substantially improves the fidelity of our extractions, which are crucial for the quantitative metrics we develop and our landscape analysis in \autoref{sec:results}; together, these constitute our primary contributions beyond \cite{huang2024analyzing} and other prior work. 

For each topic label (\texttt{purposes}, \texttt{handling}, \texttt{sharing}, and \texttt{rights}), we first concatenate all sections assigned that label (either as a primary or secondary label) and provide the resulting text to the LLM. We then instruct the LLM to extract all relevant statements and to infer both practice-specific and a common set of fields. The common fields are \texttt{related data}, \texttt{related subjects}, and \texttt{negated}. \texttt{related data} contains three boolean subfields: \texttt{global}, indicating that the statement applies to all collected data; \texttt{underspecified}, indicating that the statement applies to a subset of collected data, but the text does not clearly identify the full set of relevant data elements; and \texttt{non-identifiable}, indicating that the statement applies only to de-identified, anonymized, or aggregated data. When the statement does not apply globally, \texttt{related data} also records the relevant \texttt{categories} and \texttt{descriptors} to which the statement applies. \texttt{related subjects} records the subjects to whom the statement applies, and \texttt{negated} indicates whether it expresses a negation or an exclusion (e.g., ``we do not share \dots'').

To help the LLM infer these attributes, we supply the complete set of categories, descriptors, extracted phrases (along with the full sentence in which each phrase appears), and subjects from the data element extraction step (\autoref{sec:data-element}). Prior approaches that link related privacy policy statements, see e.g., \cite{cui2023poligraph,yang2025automated,zhao2025llm}, primarily link information within the same sentence or local context, which can miss dependencies established earlier in the text, such as definitions of collected data elements. In contrast, we treat the data collection section as global context: we first extract and consolidate all data collection statements, and then use them to ground subsequent extractions of data practices and to link each practice to the appropriate data elements, resulting in more complete and internally consistent outputs.

Practice-specific fields are defined as follows. For \texttt{purposes}, \texttt{rights} (comprising \texttt{choices} and \texttt{access}), and \texttt{handling} (comprising \texttt{retention} and \texttt{protection}), \texttt{category} records the applicable taxonomy label. For data handling statements related to retention, we additionally extract \texttt{duration}, which records the retention period if specified, \texttt{reference point}, which records the event from which the retention period is measured, and \texttt{action}, which records the action performed at the end of the retention period. For \texttt{sharing}, we extract \texttt{recipients}, which records the third-party recipients with whom data is shared, and \texttt{purposes}, which records the purposes for which the data is shared. The valid values for these fields are listed in \autoref{table:schema} in \autoref{app:tables}. Compared to \cite{huang2024analyzing}, we add extraction of data sharing statements and extend the representation of retention policies with the additional \texttt{reference point} and \texttt{action} attributes.

\section{Pipeline Evaluation}\label{sec:evaluation}
In this section, we evaluate our pipeline by reporting dataset attrition across processing stages and the results of manual validation. For all LLM outputs, we use OpenAI's \texttt{gpt-5.1} with the reasoning effort set to low.

\subsection{Evaluation Corpus}

To gather a set of domains for evaluation, we combine two sources: companies in the Russell 3000 index (also used by \cite{huang2024analyzing}), and domains from the Tranco list~\cite{pochat2018tranco}.

\paragraph{Russell 3000}

We identify Russell 3000 companies using constituents of the Vanguard Russell 3000 ETF as of 2025-10-31, yielding 2,964 company names and their Global Industry Classification Standard (GICS) sector labels across 11 sectors. We map each company to its primary domain using an LLM with search-engine results provided via tool call, yielding 2,926 unique domains after resolving redirects and removing duplicates (since some companies, e.g., Alphabet, have multiple tickers listed in the index). We then crawled these domains, with 2,276 (78\%) returning at least one candidate privacy policy page with qualifying text (as described in \autoref{sec:retrieval}). Of the 650 domains without qualifying text, 44\% are due to pages under 5,000 characters after boilerplate removal, 39\% are due to the crawler failing to identify any candidate privacy policy pages, 13\% are due to the crawler failing to reach the website's homepage (e.g., offline website or blocked by bot detection mechanisms), and 3\% are due to HTML parsing errors and non-English pages.

\paragraph{Tranco}

We use crawls of all Tranco domains from a snapshot on 2025-02-26 (4.25 million domains), crawled between 2025-02-28 and 2025-04-05. Across all Tranco domains, only 12.5\% yield at least one candidate privacy
policy page with qualifying text. This lower success rate relative to Russell 3000 companies is expected: Tranco includes many domains that do not host a conventional website (e.g., API endpoints), non-English websites, and websites that lack a privacy policy or provide only a short one (e.g., blogs). From the subset of crawls with qualifying text, we construct two evaluation samples. First, we uniformly sample domains from the top 30k (by Tranco rank) and select crawls until we obtain 3,000 domains with qualifying text. Second, we uniformly sample additional domains from the full snapshot until we obtain 4,724 domains with qualifying text.

\paragraph{Combined corpus}

Across these three strategies (Russell 3000, top-ranked Tranco, and random Tranco), we obtain 10,000 domains with qualifying text for our analysis. This yields a diverse mixture of large public companies, as well as high-traffic (but not necessarily public) and smaller entities, enabling evaluation across a broad range of organizations. We also assign each company an industry sector by prompting the LLM to infer the sector from its homepage text, using the 11 top-level GICS sectors plus ``Public Administration'' and ``Other''. Using Russell-3000 ground-truth labels, the accuracy of the LLM-generated sectors is 92.9\%.

\subsection{Evaluation Results}

\begin{table}[t]
    \caption{Summary statistics for topic-specific text and LLM extractions across our corpus (10,000 domains). Averages are computed over domains with qualifying text or extractions, respectively. Combined statistics count domains with qualifying text/extractions for at least one topic, and average total text length and total extractions across all topics.}
    \label{tab:topic-stats}
    \setlength{\tabcolsep}{3pt}
    \centering
    \footnotesize
    \begin{tabular}{lcccc}
        \hline
        \multirow{2}{*}{Topic} & \# domains & \# domains & Avg. text & Avg. \#  \\
         & with text & with extractions & length (chars) & extractions \\
        \hline
        \texttt{elements} & 9595 (95.9\%) & 9540 (95.4\%) & 13,795 & 30.3 \\
        \texttt{purposes} & 9486 (94.9\%) & 9315 (93.1\%) & 10,332 & 39.2 \\
        \texttt{handling} & 9230 (92.3\%) & 8777 (87.8\%) & 8,523 & 13.1 \\
        \texttt{sharing} & 9538 (95.4\%) & 8909 (89.1\%) & 11,267 & 13.4 \\
        \texttt{rights} & 9352 (93.5\%) & 8089 (80.9\%) & 10,551 & 12.4 \\
        \hline
        \texttt{combined} & 9659 (96.6\%) & 9556 (95.6\%) & 53,340 & 103.5 \\
        \hline
    \end{tabular}
\end{table}

\newcommand{\jkey}[1]{\textcolor{red!60!black}{\ttfamily #1}}
\newcommand{\jval}[1]{\textcolor{black}{\ttfamily #1}}

\begin{figure}[t]
    \centering
    \begin{tcolorbox}[
      title=Context from privacy policy,
      colback=white, colframe=gray!100,
      left=5pt, right=5pt, top=3pt, bottom=3pt,
      fonttitle=\footnotesize\bfseries,
    ]
        \scriptsize
        When you visit our website, we may automatically collect (and/or contract with third parties that provide) the following information:
        \begin{itemize}[leftmargin=*, noitemsep, topsep=0pt]
            \item technical information, including the IP address used to connect your computer to the Internet, browser type and version, time zone setting, \hl{browser plug-in types and versions}, operating system and platform; and
            \item \ldots
        \end{itemize}
    \end{tcolorbox}
    \begin{tcolorbox}[
      title=Structured extraction,
      colback=white, colframe=gray!100,
      left=5pt, right=5pt, top=3pt, bottom=3pt,
      fonttitle=\footnotesize\bfseries,
    ]
        \scriptsize\ttfamily
        \begin{minipage}[t]{0.4\linewidth}
            \jkey{category}: \jval{device info} \\
            \jkey{descriptor}: \jval{plugins}
        \end{minipage}\hfill
        \begin{minipage}[t]{0.56\linewidth}
            \jkey{subjects}: \jval{individual} \\
            \jkey{collectors}: \jval{first party, processor} \\
            \jkey{method}: \jval{automatic}
        \end{minipage}
    \end{tcolorbox}
    \caption{Example extraction of a data element.}
    \label{fig:element-extraction}
\end{figure}

\begin{figure}[t]
    \centering
    \begin{tcolorbox}[
      title=Context from privacy policy,
      colback=white, colframe=gray!100,
      left=5pt, right=5pt, top=3pt, bottom=3pt,
      fonttitle=\footnotesize\bfseries,
    ]
        \scriptsize
        \textbf{The Types of Information We Collect} \\
        \ldots \\
        Optional information (deliberately sent):
        \begin{itemize}[leftmargin=*, noitemsep, topsep=0pt]
          \item \emph{E-mail}: your name, e-mail address, and the content of your e-mail.
          \item \emph{Online forms}: all the data you choose to fill in or confirm. This may include credit or debit card information if you are ordering a product or making a payment, as well as information about other people if you are providing it for delivery purposes, etc.
        \end{itemize}
        \ldots \\
        \textbf{How We Use Information} \\
        \ldots \\
        Optional information enables us to \hl{provide services or information tailored more specifically to your needs}, to forward your message or inquiry to another entity that is better able to do so, and to plan Web site improvements.
    \end{tcolorbox}
    \begin{tcolorbox}[
      title=Structured extraction,
      colback=white, colframe=gray!100,
      left=5pt, right=5pt, top=3pt, bottom=3pt,
      fonttitle=\footnotesize\bfseries,
    ]
        \scriptsize\ttfamily
        \jkey{value}: purpose\,$\rightarrow$\,user experience \\
        \jkey{related data\,$\to$\,flags}: \\
        \hspace*{5pt} \jval{global=false, under-specified=false, non-identifiable=false} \\
        \jkey{related data\,$\to$\,categories}: \\
        \hspace*{5pt} \jval{personal identifier, contact Info, communication Data,} \\
        \hspace*{5pt} \jval{financial info, other} \\
        \jkey{related data\,$\to$\,descriptors}: \\
        \hspace*{5pt} \jval{name, email address, message logs, online form data,} \\
        \hspace*{5pt} \jval{payment card info, delivery contact info} \\
        \jkey{related subjects}: \jval{individual , contact}
    \end{tcolorbox}
    \caption{Example extraction of a data practice.}
    \label{fig:practice-extraction}
\end{figure}

\autoref{tab:topic-stats} summarizes extractions across our corpus. 96.6\% of the 10,000 domains contain text for at least one core topic (\texttt{elements}, \texttt{purposes}, \texttt{handling}, \texttt{sharing}, or \texttt{rights}), with an average of 103.5 extractions per domain among those with at least one extraction. A substantial share of the extraction failures in the second column of \autoref{tab:topic-stats} are due to OpenAI's content filter, with 699 domains lacking extractions for at least one topic due to the content filter.

Figures \ref{fig:element-extraction} and \ref{fig:practice-extraction} show example extractions of a collected data element and a data practice (a data collection purpose) produced by our pipeline. Note that each structured extraction corresponds only to the highlighted text; other extractions (e.g., additional data elements mentioned in \autoref{fig:element-extraction}) are captured separately. Both figures highlight the importance of providing the LLM with full context for accurate extraction. In \autoref{fig:element-extraction}, the introductory sentence to the itemized list is crucial for determining that the data are collected automatically and may be collected by either the first party or its processors. More importantly, \autoref{fig:practice-extraction} shows that relevant context may come from entirely separate sections. Here, the definition of ``optional information'' is reused to specify how that subset of collected data is used. Therefore, providing the LLM with text describing data collection (outputs of \autoref{sec:data-element}) for subsequent extractions of data practices (\autoref{sec:data-practice}) is key to accurately mapping these relationships.

\subsection{End-to-end Validation}\label{sec:validation}

To validate our system's outputs, we built a custom web interface that presents randomly selected LLM-extracted statements alongside the raw privacy policy text, while hiding the LLM-assigned attributes, to a human evaluator. Evaluators first determine whether each extraction is valid, such as whether a statement was correctly identified as data sharing. For valid extractions, they independently select the applicable attributes (e.g., data element categories/descriptors) without seeing the LLM's outputs. We validated enough samples to obtain 100 valid extractions for each of the following topics: \texttt{elements}, \texttt{purposes}, \texttt{handling} (50 \texttt{retention} and 50 \texttt{protection}), \texttt{sharing}, and \texttt{rights} (50 \texttt{choices} and 50 \texttt{access}). Evaluations were performed by two of the authors, each responsible for half of the validations for each topic. Across validations, 100/101 extracted data elements (99.0\%) and 400/405 extracted data practices (98.8\%) were valid.

After this initial round, cases where the LLM and human evaluator disagree are shown to the other evaluator, who selects the option they find more accurate without knowing which was produced by the LLM. This adjudication accounts for human error and for statements that are vague or reasonably interpreted in multiple ways. This allows us to test whether, among disagreements, either the LLM or the initial evaluator is favored at a statistically significant rate.

\begin{table}[t]
    \caption{Blind validation results. Stage 1 agreement reports the percentage of samples for which the initial human evaluator and the LLM agree. Stage 2 LLM selected reports, among stage 1 disagreements, the percentage of cases in which the second human evaluator selects the LLM output. The $p$-value tests whether the second-stage selections differ significantly from chance; statistically significant winners ($\alpha = 0.05$) are reported in the final column.}
    \label{tab:validation}
    \centering
    \scriptsize
    \setlength{\tabcolsep}{2pt}
        \begin{tabular}{llcccc}
        \hline
        \multirow{2}{*}{Topic} & \multirow{2}{*}{Field} & Stage 1 & Stage 2 & \multirow{2}{*}{$p$-value} & \multirow{2}{*}{Winner} \\
         & & agreement & LLM selected & & \\
        \hline
        \multirow{5}{*}{\texttt{elements}} & \texttt{category} & 82/100 (82.0\%) & 8/18 (44.4\%) & 0.815 & None \\
        & \texttt{descriptor} & 65/100 (65.0\%) & 15/35 (42.9\%) & 0.500 & None \\
        & \texttt{subjects} & 96/100 (96.0\%) & 3/4 (75.0\%) & 0.625 & None \\
        & \texttt{collectors} & 88/100 (88.0\%) & 5/12 (41.7\%) & 0.774 & None \\
        & \texttt{methods} & 83/100 (83.0\%) & 9/17 (52.9\%) & 1.000 & None \\

        \hline
        \multirow{1}{*}{\texttt{purposes}} & \texttt{category} & 74/100 (74.0\%) & 16/26 (61.5\%) & 0.327 & None \\

        \hline
        \multirow{3}{*}{\texttt{handling}} & 
        \texttt{category} & 80/100 (80.0\%) & 15/20 (75.0\%) & 0.041 & \cellcolor{green!20}LLM \\
        & \texttt{retention action} & 35/50 (70.0\%) & 11/15 (73.3\%) & 0.118 & None \\
        & \texttt{retention ref.} & 31/50 (62.0\%) & 13/19 (68.4\%) & 0.167 & None \\

        \hline
        \multirow{2}{*}{\texttt{sharing}} & \texttt{purposes} & 52/100 (52.0\%) & 30/48 (62.5\%) & 0.111 & None \\
        & \texttt{recipients} & 73/100 (73.0\%) & 19/27 (70.4\%) & 0.052 & None \\

        \hline
        \multirow{1}{*}{\texttt{rights}} & \texttt{category} & 80/100 (80.0\%) & 17/20 (85.0\%) & 0.003 & \cellcolor{green!20}LLM \\

        \hline
        \multirow{5}{*}{\texttt{common}} & \texttt{global} & 297/400 (74.2\%) & 53/103 (51.5\%) & 0.844 & None \\
        & \texttt{underspecified} & 268/400 (67.0\%) & 82/132 (62.1\%) & 0.007 & \cellcolor{green!20}LLM \\
        & \texttt{non-identifiable} & 394/400 (98.5\%) & 5/6 (83.3\%) & 0.219 & None \\
        & \texttt{related subjects} & 330/400 (82.5\%) & 38/70 (54.3\%) & 0.550 & None \\
        & \texttt{negated} & 389/400 (97.2\%) & 10/11 (90.9\%) & 0.012 & \cellcolor{green!20}LLM \\
        \hline
    \end{tabular}
\end{table}

Our results are reported in \autoref{tab:validation}. Each row corresponds to one field. The table reports the initial agreement rate between the human evaluator and LLM and, among disagreements, the percentage for which the second evaluator favors the LLM output. The second-to-last column reports the $p$-value from a test of whether the second evaluator's selections differ from chance. Values below 0.05 indicate a statistically significant preference for either the LLM or the human outputs; when such a difference is significant, the final column reports which source was judged more accurate.

We find that human annotations are never judged more accurate than the LLM outputs, while the LLM is the statistically significant winner in four cases. Many disagreements in the first stage arise because the relevant distinctions are inherently nuanced and subjective. For example, a statement may say that user data is shared with third parties to measure the effectiveness of advertising campaigns, which could reasonably be construed as ``advertising and marketing'', ``analytics and research'', or both. Notably, even when the initial evaluator disagreed with the LLM in the first stage, the LLM's outputs were often deemed plausible in the second stage. Overall, our results suggest that LLM outputs are on par with, and in some cases superior to, annotations produced by trained experts. Continuing this validation effort will allow us to translate nuanced cases into more precise instructions and guidelines, and, if needed, to develop manually annotated ground-truth for additional fine-tuning.

\subsection{Overview of Extracted Data}
\label{sec:statistics}

\paragraph{Data elements}

\begin{figure}[t]
    \centering
    \includegraphics[width=0.95\linewidth]{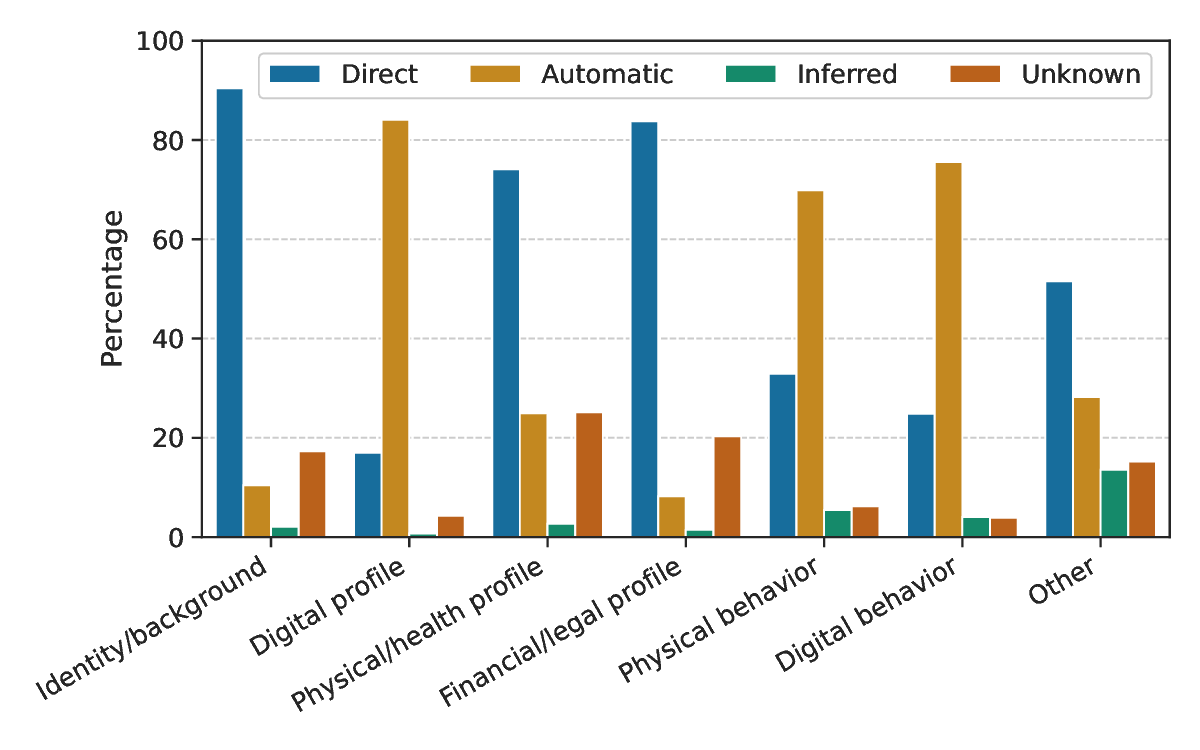}
    \caption{Percentage of collection methods among collected data elements, grouped by data meta-categories.}
    \label{fig:collection-method}
\end{figure}

A single data element (e.g., ``email address'') may be mentioned multiple times throughout a privacy policy, sometimes with respect to different data subjects or collection methods. We therefore merge duplicate annotations (with the same category and descriptor) and concatenate their associated \texttt{subjects}, \texttt{collectors}, and \texttt{methods}. After merging, the proportion of LLM-generated out-of-vocabulary data descriptors across our entire dataset is 13.0\%.

Across extracted mentions of collected data, 94.8\% concern primary individuals (i.e., the direct users/customers), and 13.5\% reference other subjects, including household members or contacts of primary individuals, personnel, job applicants, and organizational customers. The first party (the owner of the privacy policy) is identified as the initial data collector in 93.4\% of cases. In addition, 23.1\% of cases state that data may initially be collected by third parties including contracted processors, partners/affiliates, public authorities, or from public sources (e.g., social media). \autoref{fig:collection-method} also shows the proportion of data stated to be collected via different collected methods, grouped by data meta-categories. As expected, identity/background, physical profile, health data, and financial/legal data are more often collected directly, whereas digital profile and behavioral data are more often collected through automated means (e.g., browsing activity or cookies). Note that the above statistics do not necessarily sum up to one, since each data element may be associated with multiple subjects, collectors, and collection methods.

\paragraph{Data practices}

\begin{figure}[t]
    \centering
    \includegraphics[width=0.85\linewidth]{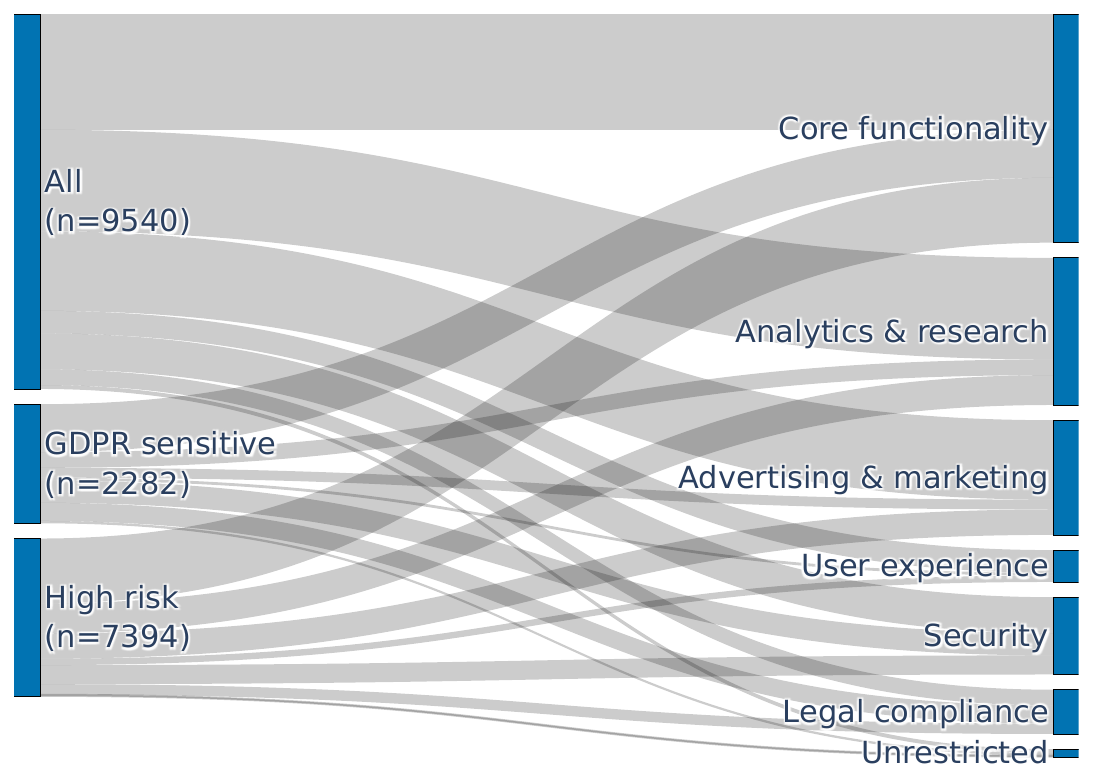}
    \caption{Proportion of companies that explicitly report sharing data with contracted partners for specific purposes.}
    \label{fig:sankey-sharing}
\end{figure}

Of all extracted data practices, 50.1\% apply globally to all collected data, and 3.6\% refer specifically to non-identifiable (anonymized/aggregated) data. Of practices tied to specific data, 56.5\% are under-specified, meaning the LLM cannot fully map them to constituent data elements.

As one way of visualizing these data practices, \autoref{fig:sankey-sharing} shows a Sankey diagram depicting how organizations report sharing data with contracted processors that perform processing on their behalf. The width of each flow reflects the percentage of companies indicating that data is shared with partners for a specific purpose. We break out these flows for (1) all data collected by a company, (2) data classified as sensitive under the GDPR, and (3) data we deem high risk (e.g., Social Security numbers).\footnote{The full list of data elements deemed GDPR-sensitive or high-risk is provided in \autoref{table:sen_schema} in \autoref{app:tables}.} We consider only scoped (non-global) statements in which the associated data are explicitly specified, and exclude negated statements and those that apply specifically to anonymized/aggregated data. For (2) and (3), we include only organizations that report collecting at least one data element from the relevant categories (2,282 for GDPR-sensitive data and 7,394 for high-risk data).

Over all data, organizations include scoped statements indicating data sharing with partners at similar rates for core functionality, analytics/research, and advertising/marketing. For high-risk data, these rates drop, more so for the latter two purposes. The drop is larger still for GDPR-sensitive data, widening the gap between core functionality and the latter two purposes. While these findings align with expectations, they also illustrate how our structured representations can surface irregular practices (e.g., using sensitive data for marketing). More importantly, beyond these targeted applications, they enable the derivation of quantitative, high-level metrics that characterize a privacy policy and support intra- and inter-sector comparisons, as we show in the next section.

\section{A Quantitative Analysis Framework}
\label{sec:analysis}
A primary motivation for making annotation automated and scalable is to obtain large-scale datasets that can enable downstream quantitative analysis, which would be otherwise infeasible. Toward this end, beyond the basic statistics such as those presented in \autoref{sec:statistics}, we are also interested in a deeper understanding of what is not easily gleaned from first-order distributional information. For instance, is a lengthy (or seemingly more complete or comprehensive) policy more transparent, or better for user protection? Is there a way to define and measure what a ``good'' privacy policy is? Setting aside compliance requirements, what purpose does such a policy generally advance, a firm's corporate and business interests or the user/consumer's protection?

In this section, we present a quantitative framework that translates our structured extractions into four metrics, each corresponding to a key \emph{dimension} of a privacy policy: (1) Completeness, (2) Transparency, (3) Business Interest, and (4) User Protection. Together, these measures enable quantitative, comparative analysis of privacy policies, and yield key insights into the privacy policy landscape, as we demonstrate in the next section. 
By explicitly distinguishing these four dimensions, we obtain a simple yet faithful high-level representation of a privacy policy.

\subsection{Metric definitions}

We now detail the formal definition of our metrics, with an example computation based on a simplified illustrative privacy policy provided in \autoref{app:metric_computation}. We represent a policy as $\mathcal{P} = (D, P, \Phi)$. Here, $D = \{d_1, \ldots, d_n\} \subseteq \mathcal{D}$ denotes the extracted data elements (where $\mathcal{D}$ is the universe of possible elements). A data element $d \in D$ consists of a category-descriptor pair (unique within $D$) along with its associated subjects, collectors, and collection methods (see \autoref{fig:element-extraction} for an example), aggregated across all mentions of the category-descriptor pair. $P=\{p_{\text{access}},p_{\text{choice}},p_{\text{retention}},p_{\text{protection}},p_{\text{purpose}},p_{\text{sharing}} \}$ denotes the six practices in our taxonomy, detailed in \autoref{sec:data-practice}. For each practice $p \in P$, let $\mathcal{V}_p$ denote its set of possible values. We formalize the policy's content as a mapping $\Phi: D \times P \rightarrow \bigcup_{p \in P} \mathcal{V}_p \cup \{\mathrm{null}\}$, such that $\Phi(d, p) \in \mathcal{V}_p \cup \{\mathrm{null}\}$ assigns
a value to practice $p$ for data element $d$, where $\mathrm{null}$ indicates that the policy does not specify practice $p$ for $d$. For instance, in \autoref{fig:practice-extraction}, ``user experience'' ($\in \mathcal{V}_{p_{\text{purpose}}}$) is assigned as the purpose for related data.

As a preprocessing step, we retain only data elements and associated practices that pertain to primary individuals who directly interact with the organization (i.e., end users/customers). While the same methodology could be applied to other populations (e.g., employees), we focus on direct users/customers, which constitute the majority of our extractions. We also ignore practices that are explicitly stated to apply only to non-identifiable data.

\paragraph{Completeness}

According to GDPR~\cite{gdpr2016}, a structurally complete policy should clearly explain what data is collected, how it is obtained, and why and how it is processed, which we distill into the Completeness score, denoted by $C$. $C$ is formulated as a weighted linear combination of three sub-metrics: (1) practice coverage ($C_{\text{practice}}$), (2) descriptor specificity ($C_{\text{desc}}$), and (3) collection method specificity ($C_{\text{method}}$):
$C = \boldsymbol{\lambda}_c^\top \cdot [C_{\text{practice}}; C_{\text{desc}}; C_{\text{method}}]$, where $\boldsymbol{\lambda}_c \in \mathbb{R}^3$ is a vector of weights. Here, $C_{\text{practice}}$ captures the extent to which the privacy policy specifies practices (across the six practices described above) for each collected data element: $C_{\text{practice}} = \frac{1}{|D|} \sum_{d \in D} \left(\frac{1}{|P|} \sum_{p \in P} \mathbf{1}\{\Phi(d, p) \neq \mathrm{null}\}\right)$, where $\mathbf{1}\{\cdot\}$ is the indicator function. $C_{\text{desc}}$ measures the proportion of mentioned data categories that include at least one specific descriptor, rather than remaining vague at the category level. Finally, $C_{\text{method}}$ measures the proportion of collected data elements for which a collection method is specified (i.e., direct, automatic, or inferred). Crucially, a privacy policy achieves a perfect Completeness score ($C=1$) if and only if it specifies all six practices for every collected data element (either through global or scoped statements), includes at least one specific descriptor for each mentioned data category, and provides an explicit collection method for each data element.

\paragraph{Transparency} Transparency measures clarity and specificity by assessing whether statements are scoped to specific data elements (as opposed to being stated globally or left unbound), and whether practice descriptions are precise rather than vague or generalized. Let $S$ denote the set of statements of data practices extracted from a privacy policy. We formulate the Transparency score, denoted as $T$, as a weighted linear combination of scope specificity ($T_{\text{scope}}$) and linguistic clarity ($T_{\text{clarity}}$): $T = \boldsymbol{\lambda}_t^T \cdot [T_{\text{scope}}; T_{\text{clarity}}]$, with weights $\boldsymbol{\lambda}_t \in \mathbb{R}^2$. Here $T_{\text{scope}} = \frac{1}{|S|} \sum_{s \in S} w_{\text{scope}}(s)$ measures, on average, how narrowly statements are scoped to collected data, with $w_{\text{scope}} \in \{w_\text{global}, w_\text{underspec}, w_{\text{spec}}\}$. For example, the practice in \autoref{fig:practice-extraction} receives $w_{\text{spec}}$, it would receive $w_\text{underspec}$ if the under-specified flag were set, and $w_\text{global}$ if the global flag were set. Additionally, $T_{\text{clarity}} = \frac{1}{|S|} \sum_{s \in S} w_{\text{clarity}}(s)$ captures ambiguity in wording, where $w_{\text{clarity}}(s) \in \{w_{\text{vague}}, w_{\text{clear}}\}$ assigns a weight depending on whether $s$ is considered vague or clear. We treat the following types of statements/phrases as vague: an unrestricted collection purpose, a limited but unspecified retention period, a generic mention of data protection, or sharing whose purpose is unrestricted or whose recipient is an unspecified third party.

\paragraph{Business Interest and User Protection} 

Business Interest and User Protection scores characterize the extent to which a privacy policy is oriented toward each of these factors, respectively. In our schema, Business Interest is determined by the stated purposes of data collection and processing, as well as data sharing practices, which reflect an organization's intention to monetize data or otherwise extract value from it. In contrast, User Protection is measured based on data handling (comprising data retention and data protection) and user rights (comprising user choices and user access) statements. These practices can reduce risks and protect/empower users by implementing institutional safeguards and policies and by providing mechanisms for users to control the collection and processing of their data, as well as to access or modify it. Note that the goal of these scores, especially Business Interest, is not to characterize certain practices as negative or non-compliant, but to quantify how the privacy policy balances these two potentially competing interests. Formally, we define these scores as weighted linear combinations of practice-specific sub-scores: $B = \boldsymbol{\lambda}_b^\top \cdot [Q_{\text{purpose}}; Q_{\text{sharing}}], U = \boldsymbol{\lambda}_u^\top \cdot [Q_{\text{retention}}; Q_{\text{protection}}; Q_{\text{choice}}; Q_{\text{access}}]$.
To compute the underlying sub-score $Q_p$ for any given practice $p \in P$, we account for both the importance of each declared practice value to the relevant factor and the breadth of its applicability across collected data: $Q_p = \sum_{v_p \in \mathcal{V}_p} w_{v_p} \times F(v_p)$. Here, $F(v_p)$ represents the fraction of collected data elements governed by value $v_p$: $F(v_p) = \frac{1}{|D|} \sum_{d \in D} \mathbb{I}\{\Phi(d, p) = v_p\}$; this accounts for how broadly a practice applies across collected data, distinguishing widely applicable practices from those limited to a narrow subset of the data. Finally, $w_{v_p} \in [0,1]$ denotes an expert-defined weight reflecting the degree of User Protection or Business Interest associated with a given value. For example, detailed privacy settings are assumed to provide users with greater control than opt-out mechanisms that require directly contacting the organization, such as by email or mail, and can therefore be assigned a higher weight.

\subsection{Parameter Selection}

Our framework relies on a number of parameters: mixture weights for aggregating sub-metrics within each dimension, weights used to define the Transparency score, and weights assigned to different practice values for computing the Business Interest and User Protection scores. We now describe how we set these weights for the remainder of our analysis.

We set the mixture weights $(\boldsymbol{\lambda}_c, \boldsymbol{\lambda}_t, \boldsymbol{\lambda}_b, \boldsymbol{\lambda}_u)$ in a data-driven manner by applying principal component analysis (PCA) to the sub-metrics and taking the loading vector of the first principal component. The intuition is that sub-metrics within a dimension (e.g., $T_{\text{scope}}$ and $T_{\text{clarity}}$ for Transparency) should reflect related aspects of a common latent factor. PCA identifies the linear combination that captures the greatest shared variance among these sub-metrics, producing a composite score that minimizes information loss while preserving the dominant direction along which they co-vary. We then apply a winsorized min-max normalization to the resulting composite scores, clipping extreme values at fixed lower and upper percentiles before rescaling to the interval $[0,1]$.

To set the practice-value weights $w_{v_p}$ used in the Business Interest and User Protection scores, we use two schemes: expert-assigned weights and survey-derived weights. For the expert weights, we use our domain expertise to group relevant practice values into tiers according to their perceived contribution to Business Interest or User Protection, and assign weights in $[0,1]$ based on these rankings. For the survey-derived weights, we recruit 50 participants on Prolific~\cite{prolific} with expertise in cybersecurity, data privacy, privacy law, or data science. Participants rate each relevant practice value on a 1--5 importance scale; we normalize ratings to $[0,1]$ and use their mean as the survey-derived weight. We further compute 95\% confidence intervals using non-parametric bootstrapping of the mean. Both sets of weights are reported in \autoref{table:schema_weights} in \autoref{app:tables}.

To compare the two sets of weights, we compute Business Interest and User Protection scores for all privacy policies in our dataset under both expert-assigned and survey-derived weights, then measure the Spearman rank correlation between the resulting rankings. The correlations are 98.1\% for Business Interest and 99.4\% for User Protection. As an additional sensitivity analysis, we repeatedly sample alternative survey-derived weight vectors by drawing each weight from its bootstrapped confidence interval and recompute rankings. Across 100 repetitions, the mean rank correlations are 98.1\%$\pm$0.12\% for Business Interest and 99.4\%$\pm$0.04\% for User Protection, where $\pm$ denotes standard deviation. These results indicate that policy rankings are highly robust to the choice of weights. Accordingly, we present the remainder of our analysis using our own expert weights.

Finally, the weights used for the Transparency score are as follows: $w_{\text{global}} = w_{\text{vague}} = 0.2$, $w_{\text{underspec}} = 0.5$, and $w_{\text{spec}} = w_{\text{clear}} = 1$. To validate these choices, we present survey participants with 13 pairs of statements. Each pair contrasts two statements that differ along one axis: vague versus clear wording (4 pairs), global versus non-global but under-specified scope (4 pairs), or under-specified versus fully scoped statements (5 pairs). Participants are asked to select the statement they perceive as more transparent. In 9 out of 13 cases, participants showed a statistically significant preference for one statement ($p<0.05$); 8 of these 9 preferences aligned with our expected ordering, with the only exception involving a global versus underspecified comparison. 2 of the 4 non-significant cases also involved global versus underspecified statements, suggesting that this distinction is the most ambiguous. Notably, participants largely agreed with the ordinal structure imposed by our Transparency weights despite receiving intentionally broad instructions, being asked to judge transparency without specified criteria, and seeing only the raw statements rather than our structured representations. The pairs for which participants did not reach consensus, as well as the one pair for which the consensus contradicted our expectation, are provided in \autoref{tab:disagreements} in \autoref{app:tables}. \autoref{app:sensitivity} presents a sensitivity analysis of the Transparency score with respect to the choice of weights.

\section{Detailed Analysis and Results}\label{sec:results}

In this section, we analyze the privacy policy landscape and sector differences using the previously defined metrics.

\subsection{Sector Differences} 

\begin{figure}
    \centering
    \includegraphics[width=1\linewidth]{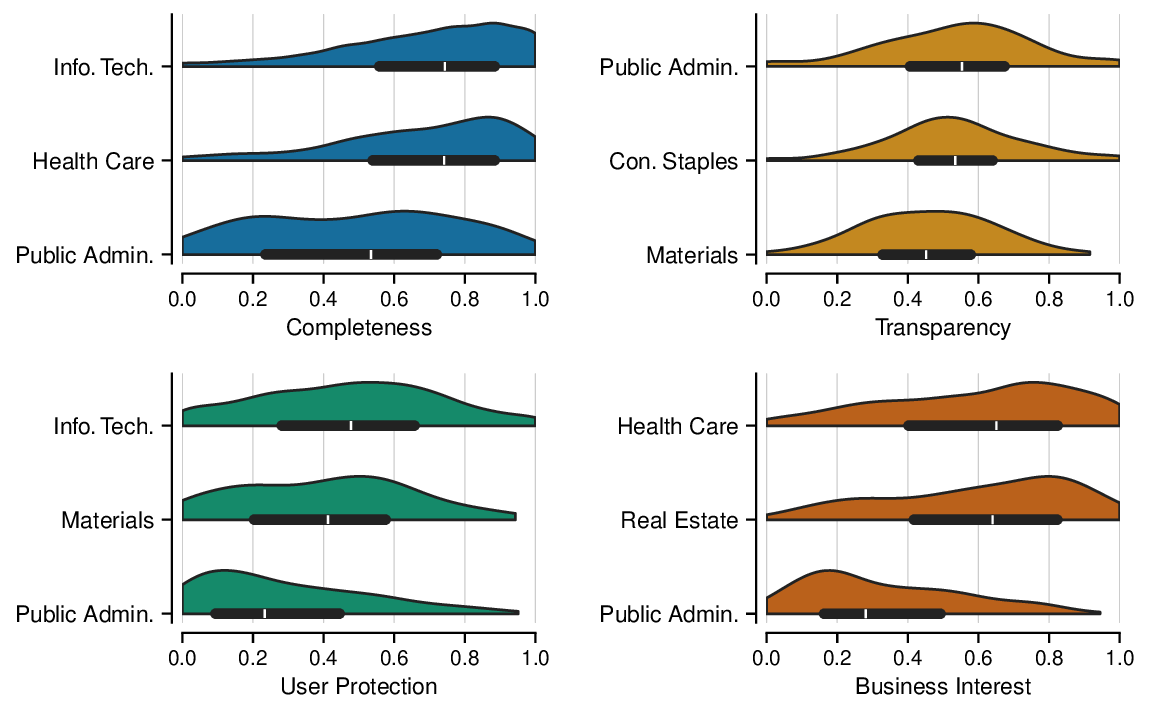}
    \caption{Sector-wise distribution of privacy policy metrics. The inner box spans the first and third quartiles, with the white line indicating the median; a kernel density estimate (KDE) overlays the box to visualize the distribution (Info. Tech.: Information Technology, Comm.: Communications, Con. Staples: Consumer Staples, Con. Disc.: Consumer Discretionary, Public Admin.: Public Administration).}
    \label{fig:sector_distribution_landscape}
\end{figure}

\autoref{fig:sector_distribution_landscape} shows the distribution of our four metrics for select sectors. This is a violin plot, consisting of a box plot (with edges indicating the first and third quartiles, and white line marking the median) and a kernel density estimate (KDE). For each metric, we show the two sectors with the highest and the one with the lowest median (excluding ``other'').

Transparency shows the highest concentration across sectors, with slight differences in sector medians. In contrast, other metrics exhibit greater variability (e.g., Public Administration displays a wide spread in Completeness). Notably, Public Administration shows a distinctive profile: although its policies score relatively high on Transparency, they perform poorly on other metrics. This combination suggests that public-sector privacy policies may emphasize clarity while providing limited details on data practices and safeguards. The resulting imbalance points to a design that prioritizes readability or compliance over coverage.

Moreover, although Information Technology and Health Care lead in Completeness, Information Technology scores attains the highest median score for User Protection, whereas Health Care scores highest on Business Interest. A closer look shows that Health Care organizations report the highest average number of unique purposes, both for internal use and sharing, in their privacy policies, followed by Real Estate, which has the second-highest Business Interest score. Notably, Health Care and Real Estate also report the highest rates of ``unrestricted'' data sharing, an observation that is especially surprising for Health Care. In contrast, Information Technology firms cite the highest average number of protection mechanisms and user access choices.

\subsection{Treatment of Sensitive Data}

\begin{figure}
    \centering
    \includegraphics[width=1\linewidth]{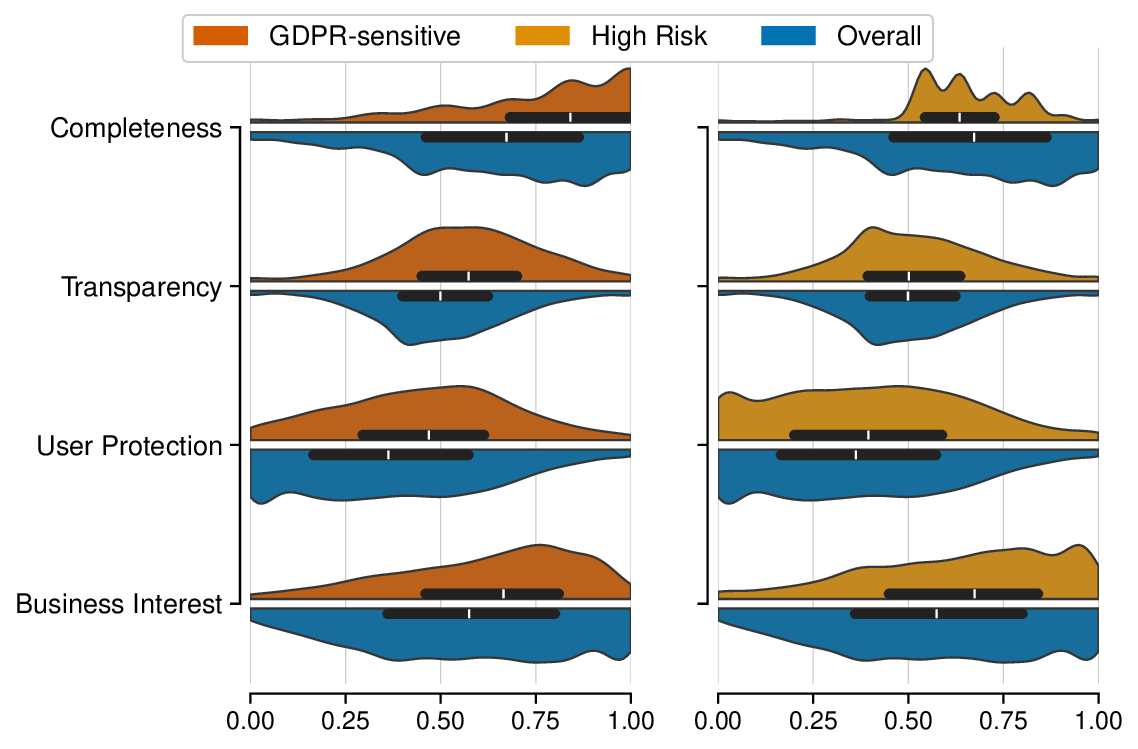}
    \caption{Comparison of metric distributions computed over all data, GDPR-sensitive data, and high-risk data.}
    \label{fig:sensitivity_split_violin}
\end{figure}

To further examine heterogeneity in how privacy policies treat sensitive versus non-sensitive data across sectors, we repeat our analysis separately for two categories of sensitive data (also used in \autoref{sec:statistics}): data defined as sensitive under the GDPR, and our own classification of high-risk data (e.g., financial data or sensitive identifiers including Social Security numbers); the full list of data elements deemed GDPR-sensitive or high-risk is provided in \autoref{table:sen_schema} in \autoref{app:tables}. We compute each of our four metrics \emph{only} for data belonging to these categories and their governing data practices. This allows us to assess whether privacy policies tend to be more complete or transparent with respect to sensitive data, and whether stronger protections or business-oriented practices are selectively applied to such data. We exclude policies that do not mention collecting any GDPR-sensitive or high-risk data from their corresponding analyses, respectively.

\autoref{fig:sensitivity_split_violin} shows a comparison between the overall metric values computed using all data and the metrics computed only for GDPR-sensitive and high-risk data. We observe an upward shift in Completeness and Transparency for GDPR-sensitive data. In contrast, for high-risk data, Transparency is largely unchanged and Completeness is slightly lower. This is likely due to the effects of regulation: GDPR-sensitive data are explicitly defined and subject to heightened scrutiny, which forces more complete and transparent disclosures. We also observe stronger user protections for GDPR-sensitive data and, to a lesser extent, for high-risk data. Interestingly, both also exhibit higher Business Interest scores, potentially an indication that these sensitive data types are closely associated with corporate business models.

\begin{figure}[t]
    \centering
    \includegraphics[width=1\linewidth]{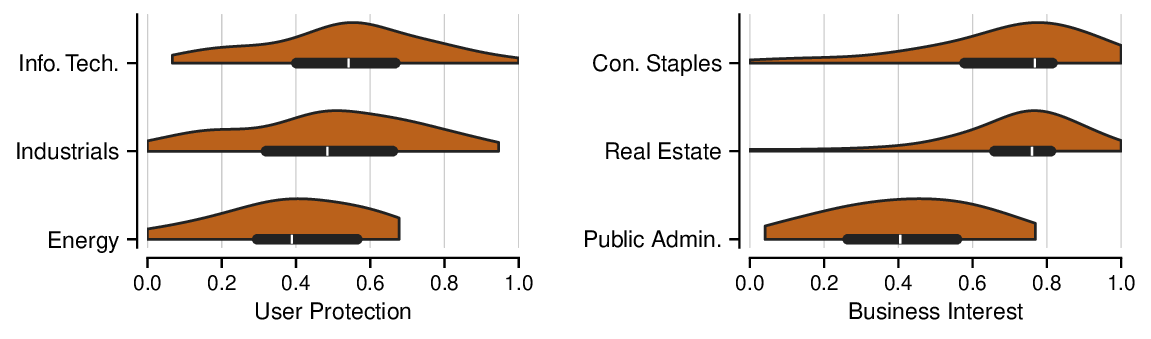}
    \caption{Sector-wise distributions of User Protection and Business Interest scores for GDPR-sensitive data.}
    \label{fig:sen-sector-violin}
\end{figure}

\autoref{fig:sen-sector-violin} also shows sector-wise distributions for GDPR-sensitive data, where similar to \autoref{fig:sector_distribution_landscape} we show the distribution of our metrics for the two highest scoring sectors and the lowest scoring sector by median score. Here, we only show distributions for User Protection and Business Interest. Information Technology retains the highest median score on User Protection, while Public Administration retains the lowest on Business Interest. However, Consumer Staples replaces Health Care (now ranking 5th) as the top sector on Business Interest, suggesting that Health Care is comparatively more cautious about monetizing sensitive data, possibly reflecting stricter regulatory requirements for medical information.

\subsection{Completeness vs. Transparency}

\begin{figure}
    \centering
    \begin{subfigure}[t]{0.49\linewidth}
        \includegraphics[width=\linewidth]{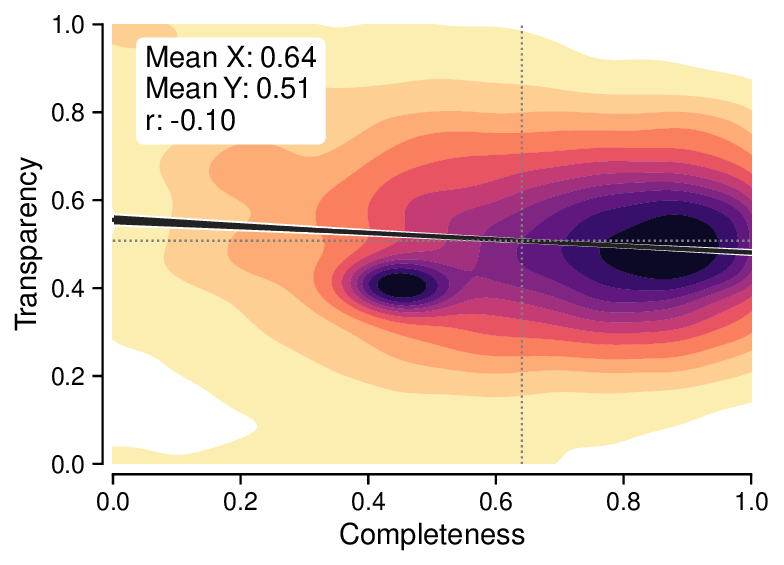}
        \captionsetup{justification=centering}
        \caption{Completeness vs.\\ Transparency}
        \label{fig:global_landscape_kde_completeness_transparency}
    \end{subfigure}\hfill
    \begin{subfigure}[t]{0.49\linewidth}
        \includegraphics[width=\linewidth]{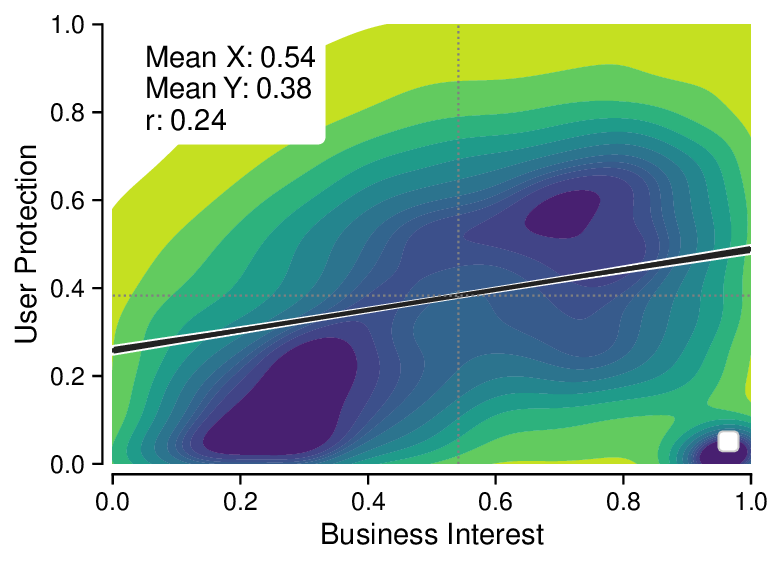}
        \captionsetup{justification=centering}
        \caption{User Protection vs.\\ Business Interest}
        \label{fig:global_landscape_kde_org_user}
    \end{subfigure}
    \caption{Joint distributions between metric pairs. Shaded KDE contours show density, with darker regions indicating higher concentration. We also overlay an ordinary least squares regression line and quantify the direction and strength of association using the Pearson correlation coefficient ($r$).}
\end{figure}

We next examine the relationship between Completeness and Transparency. \autoref{fig:global_landscape_kde_completeness_transparency} visualizes their joint distributions using KDE, where darker regions indicate higher concentration. We also overlay a trend line (fitted via ordinary least squares linear regression) to characterize the direction and strength of association. We observe a weak negative correlation between the two (Pearson correlation coefficient $r=-0.1$). This suggests that Completeness and Transparency are largely independent, meaning that how complete or lengthy a privacy policy is has all but nothing to do with how transparent it is, and if anything, more complete/lengthier policies tend to be less transparent. We also observe two concentration regions along the completeness axis, but none along Transparency (consistent with the distributions in \autoref{fig:sensitivity_split_violin}). 

\subsection{User Protection vs. Business Interest}

\begin{figure}[t]
    \centering
    \includegraphics[width=0.9\linewidth]{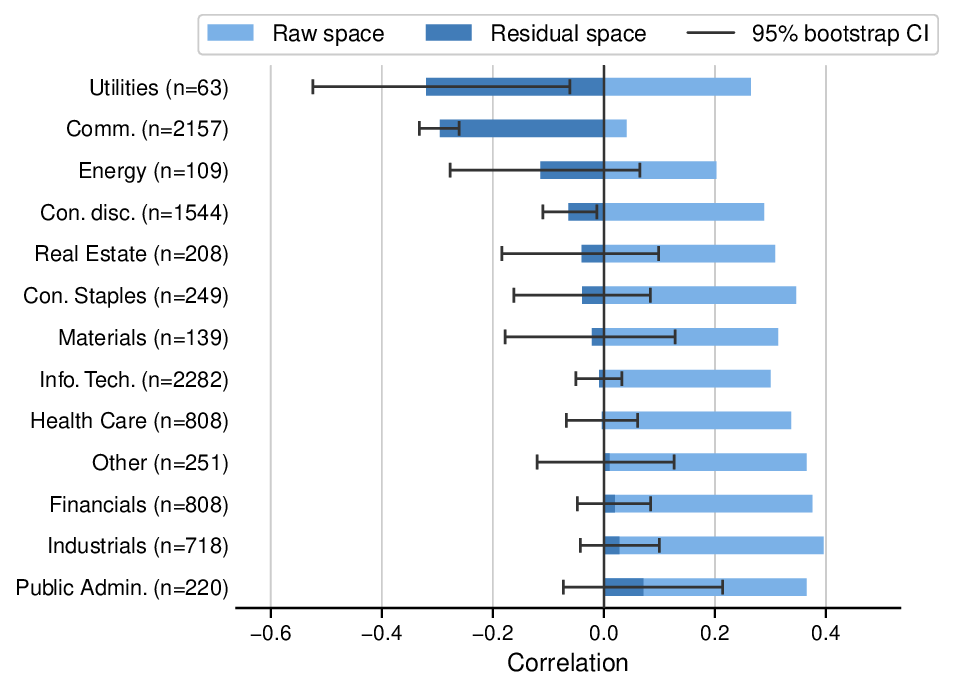}
    \caption{Sector-wise correlations between User Protection and Business Interest.} 
    \label{fig:sector_raw_vs_residual_correlation}
\end{figure}

Similarly, \autoref{fig:global_landscape_kde_org_user} shows the joint distribution of User Protection and Business Interest. At first glance, we observe a moderate positive association: organizations with more business-oriented privacy policies also tend to score higher on user protection. However, completeness is strongly associated with Business Interest ($r=0.49$) and User Protection ($r=0.64$); Transparency's associations are moderate to weak ($r=-0.21$ and $r=-0.05$, respectively). This leads us to control for confounding: we fit a regression model with Completeness and Transparency as covariates and use the resulting residuals to remove their effects from both scores, after which the association becomes weakly negative in the residual space ($r=-0.10$), indicating that User Protection and Business Interest are also largely independent after accounting for Completeness and Transparency; this further supports treating them as distinct metrics. 

However, the sector-level view, depicted in \autoref{fig:sector_raw_vs_residual_correlation}, is more revealing. Specifically, it reports the Pearson correlation in both raw and residual space, as described above. Notably, while the global association is weak, the sector-level decomposition reveals substantial heterogeneity. We use bootstrapping to compute 95\% confidence intervals (CIs); CIs that include zero cannot reject the null hypothesis of independence. For all but two sectors (Communications and Utilities) we cannot reject independence. Public Administration shows the largest positive association, but its CI still includes zero, likely due to a relatively small sample size. In contrast, Communications shows a moderately strong negative association, with a weaker but still negative relationship in Utilities. Communications includes large firms such as Google, Meta, AT\&T, for which commercial incentives to monetize data may more directly conflict with user protection. Interestingly, although Communications and Information Technology are often grouped together as ``tech'', they exhibit strikingly different patterns. In Information Technology, home to firms such as Microsoft, Intel, Nvidia, we find little discernible association between User Protection and Business Interest, whereas the relationship in Communications is substantially more negative.

\begin{figure}[t]
    \centering
    \includegraphics[width=0.8\linewidth]{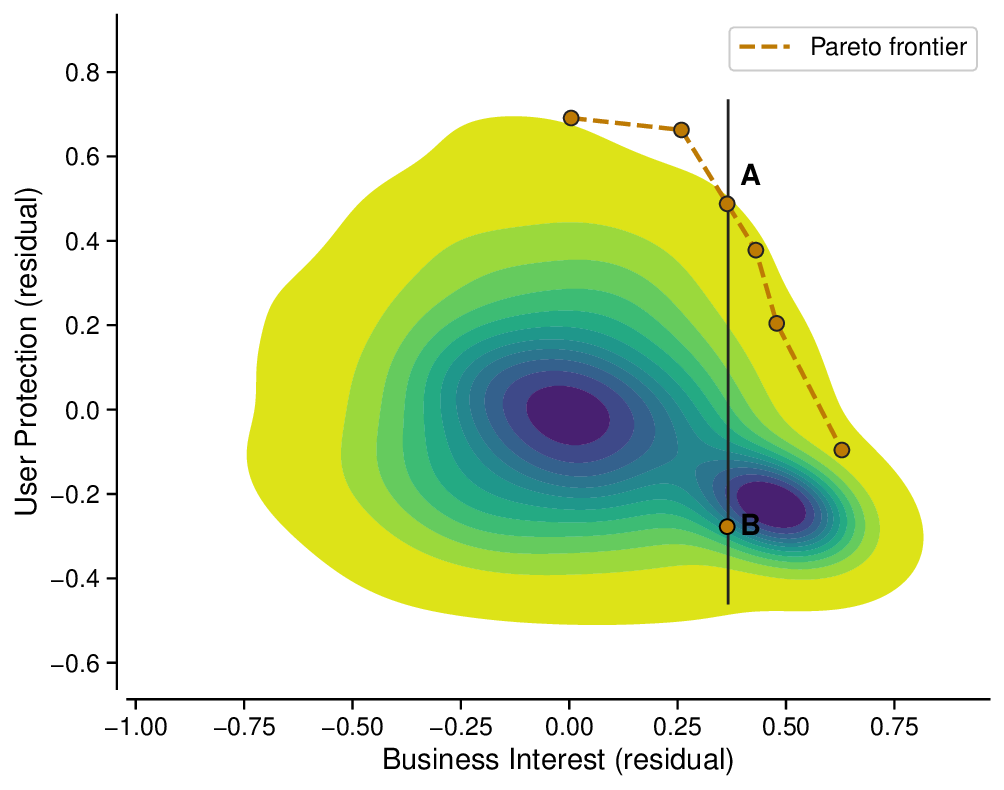}
    \caption{Residual trade-off structure with Pareto frontier between Business Interest and User Protection within the Communications sector.}
    \label{fig:pareto_frontier_communication_services}
\end{figure}

\begin{figure}[t]
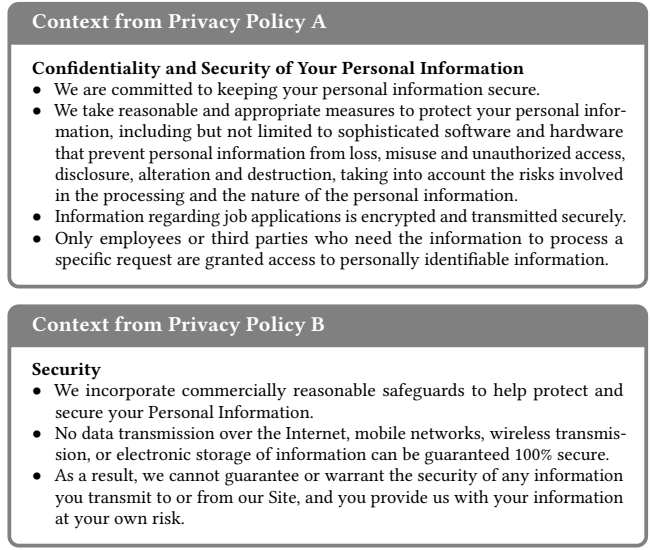

    \centering

    \begin{tcolorbox}[
      title=Context from Privacy Policy A,
      colback=white, colframe=gray!100,
      left=5pt, right=5pt, top=3pt, bottom=3pt,
      fonttitle=\footnotesize\bfseries,
    ]
        \scriptsize
        \textbf{Confidentiality and Security of Your Personal Information}
        \begin{itemize}[leftmargin=*, noitemsep, topsep=0pt]
            \item We are committed to keeping your personal information secure.
            \item We take reasonable and appropriate measures to protect your personal information, including but not limited to sophisticated software and hardware that prevent personal information from loss, misuse and unauthorized access, disclosure, alteration and destruction, taking into account the risks involved in the processing and the nature of the personal information.
            \item Information regarding job applications is encrypted and transmitted securely.
            \item Only employees or third parties who need the information to process a specific request are granted access to personally identifiable information.
        \end{itemize}
    \end{tcolorbox}

    \begin{tcolorbox}[
      title=Context from Privacy Policy B,
      colback=white, colframe=gray!100,
      left=5pt, right=5pt, top=3pt, bottom=3pt,
      fonttitle=\footnotesize\bfseries,
    ]
        \scriptsize
        \textbf{Security}
        \begin{itemize}[leftmargin=*, noitemsep, topsep=0pt]
            \item We incorporate commercially reasonable safeguards to help protect and secure your Personal Information.
            \item No data transmission over the Internet, mobile networks, wireless transmission, or electronic storage of information can be guaranteed 100\% secure.
            \item As a result, we cannot guarantee or warrant the security of any information you transmit to or from our Site, and you provide us with your information at your own risk.
        \end{itemize}
    \end{tcolorbox}
    
    \caption{Data protection statements from privacy policies A and B (\autoref{fig:pareto_frontier_communication_services}).}
    \label{fig:protection-context}
\end{figure}

Given this negative association in Communications, we next examine whether sectors that exhibit a negative association between Business Interest and User Protection (e.g., Communications) exhibit a meaningful trade-off (i.e., an implied optimal operating point or ``frontier'') against which firms within the sector can be compared. \autoref{fig:pareto_frontier_communication_services} plots this frontier in the residual space. Both the KDE and the frontier's downward slope clearly indicate the negative association between the two. The knee point, in turn, suggests a potential optimal operating point, one that offers meaningful benefits to the firm while still providing protection to users.

To illustrate, we randomly choose two firms (marked A and B in \autoref{fig:pareto_frontier_communication_services}) -- the two have near-identical Business Interest scores, but A is much higher on User Protection -- and compare their privacy policy texts. \autoref{fig:protection-context} highlights clear divergences in their data protection clauses. Whereas firm A provides explicit and concrete statements regarding the encryption of information transferred during job application processes, as well as clear restrictions on personnel authorized to access personal data, firm B does not describe any specific safeguards for information transferred to its site. Additionally, inspection of the two firms' data retention statements (included in \autoref{app:statemet_example}) reveals explicit retention periods for firm A (five years for personal information related to its services and three years for other relevant personal information). Firm B, however, does not provide any specific retention durations for personal information.

These comparisons indicate that for at least a subset of firms, it is possible to strengthen user protection provisions without compromising business interests, highlighting meaningful room for improvement in privacy policy practices beyond what is strictly required for business objectives. This also highlights the ability of our quantitative framework to identify such cases at scale.

\subsection{A Case Study: TikTok}\label{sec:tiktok}

\begin{figure}[t]
    \centering
    \begin{subfigure}[t]{0.49\linewidth}
        \includegraphics[width=\linewidth]{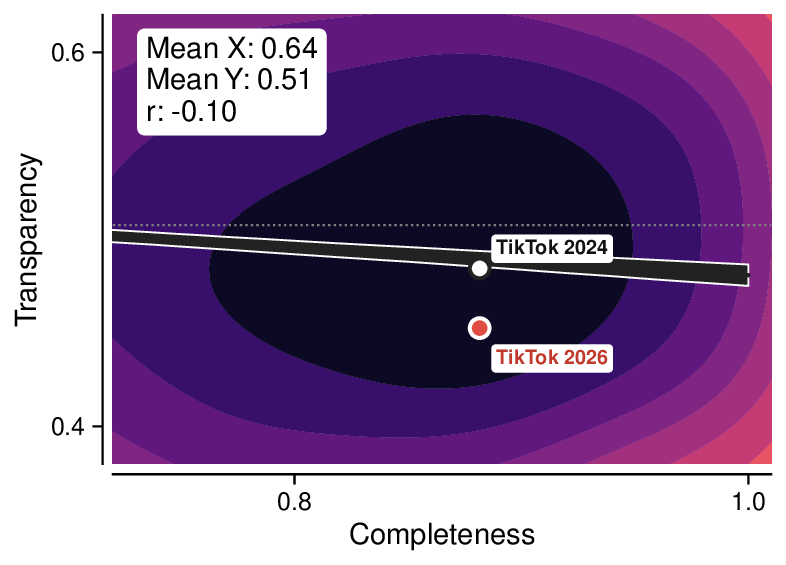}
        \captionsetup{justification=centering}
        \caption{Completeness vs. Transparency}
        \label{fig:global_landscape_kde_completeness_transparency_all}
    \end{subfigure}\hfill
    \begin{subfigure}[t]{0.49\linewidth}
        \includegraphics[width=\linewidth]{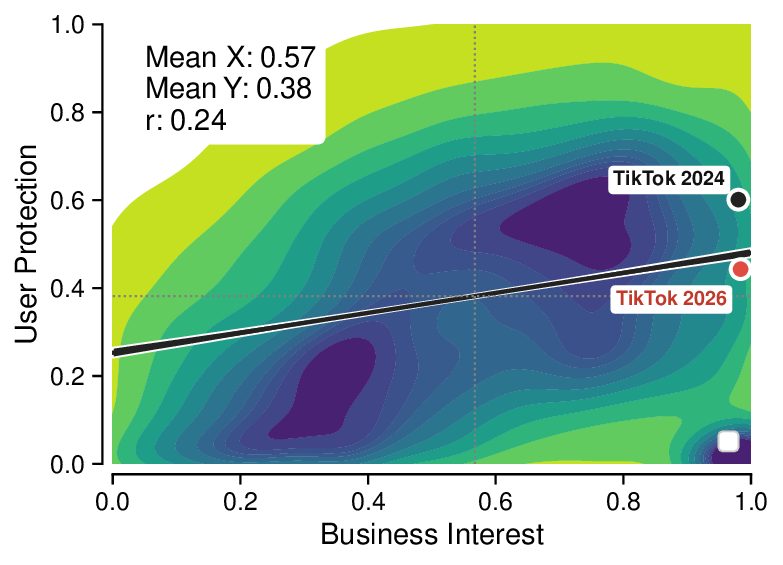}
        \captionsetup{justification=centering}
        \caption{User Protection vs. Business Interest}
        \label{fig:global_landscape_kde_org_user_all}
    \end{subfigure}
    \caption{Joint distributions between metrics pairs, with TikTok's 2024 and 2026 privacy policies overlaid in black and red, respectively.}
    \label{fig:tiktok_sensi_analysis}
\end{figure}

In early 2026, TikTok released an updated privacy policy following its change in ownership, a move that drew substantial public scrutiny. As noted by \cite{fiesler2025tiktok}, while public reactions may occasionally be amplified by misinterpretations of complex legal phrasing, the widespread inclination to assume worst-case scenarios reflects a systemic fragility in user trust toward the broader technology industry. To evaluate whether these public concerns align with changes in the policy text, we apply our quantitative evaluation framework to compare TikTok's 2024 and 2026 privacy policies. Figure \ref{fig:tiktok_sensi_analysis} overlays the scores assigned to TikTok's 2024 and 2026 policies on the joint distributions of Completeness versus Transparency and User Protection versus Business Interest, providing a longitudinal view of these changes.

We observe an uneven shift in TikTok's metrics. While the 2026 version maintains high scores in Completeness and Business Interest, it shows a clear decline in both Transparency and User Protection compared to the 2024 version. Further investigation reveals that this shift is primarily driven by a change in how the policy describes internal access to user data. Specifically, the 2024 policy limited corporate-group entities' remote access to user data to an ``as needed'' basis, whereas this limiting language was removed in the 2026 revision. This change may provide greater operational flexibility, but it also weakens an explicit safeguard against unnecessary access to user data. This further illustrates the tension between corporate incentives and user protection, as well as the choices firms make in balancing these priorities.

\section{Conclusion}
\label{sec:conclusion}
In this paper, we present an end-to-end system that converts privacy policies into fine-grained structured representations and leverages these representations to define systematic measures of completeness, transparency, and orientation toward user protection versus business interests. Our framework enables consistent, repeatable assessment of privacy policies and supports comparisons across firms and industry sectors, yielding high-level insights into the privacy policy landscape. Future work could further refine and extend the framework, for example, by evaluating jurisdiction-specific privacy policies. Improving scalability by using outputs from state-of-the-art frontier models to distill their capabilities into smaller local models, and further fine-tuning with manually curated labels, are also important aspects of our ongoing work. Finally, longitudinal analysis of privacy policies can help identify shifts in the landscape, particularly following the introduction of major regulations.

\appendix
\section{Ethics}

This study uses publicly available web content collected via web crawling as its preliminary raw data. Web crawling has long been a standard method in Internet research, and we limit collection to the minimum information necessary for our analysis. Specifically, we only navigate to each organization's privacy policy page; no practical alternative enables collection at the scale required for this study. All data collection is rate-limited and restricted to information available on the public web.

To motivate our choice of parameters, we also elicit input from cybersecurity and data privacy experts through an anonymous online survey. The survey collects only participants’ areas of expertise and their opinions on a set of privacy policy practices and statements; we do not collect any sensitive or personally identifying information.

\bibliographystyle{ACM-Reference-Format}
\bibliography{references}

\newpage
\section{Additional Tables}\label{app:tables}

\begin{center}
    \footnotesize
    \setlength{\tabcolsep}{2pt}
    \captionof{table}{Valid values for fields in our extraction schema.}
    \label{table:schema}
    \begin{tabular}{@{} llp{0.55\linewidth} @{}}
    \hline
    Topic & Field & Valid values \\
    \hline
    
    \texttt{elements} & \texttt{subjects} & individual, household member, contact, personnel, applicant, organization, other. \\
    & \texttt{collectors} & first party, affiliate, successor, processor, partner, public authority, public, other. \\
    & \texttt{methods} & direct, indirect, inferred, unknown. \\
    \hline
    
    \texttt{purposes} & \texttt{category} & core functionality, user experience, security, analytics and research, advertising and marketing, data sale, legal compliance, unrestricted, other. \\
    \hline
    
    \texttt{handling} & \texttt{protection cat.} & secure authentication, secure transfer, secure storage, access limit, privacy program, privacy review, generic. \\
    & \texttt{retention cat.} & indefinitely, limited, stated. \\
    & \texttt{retention ref.} & data collection, data de-identification, purpose completion, legal requirement expiry, account deletion, account inactivity, consent withdrawal, other. \\
    & \texttt{retention action} & deletion, de-identification, other. \\
    \hline

    \texttt{sharing} & \texttt{purposes} & Same as \texttt{category} $\rightarrow$ \texttt{purposes}. \\
    & \texttt{recipients} & Same as \texttt{elements} $\rightarrow$ \texttt{collectors}, excluding ``first party''. \\
    \hline
    
    \texttt{rights} & \texttt{choice category} & no choice, opt-in, opt-out via link, opt-out via direct contact, privacy settings. \\
    & \texttt{access category} & none, view, export, edit, restrict, deactivate, partial delete, full delete. \\
    
    \hline
    \end{tabular}
\end{center}

\begin{center}
    \captionof{table}{Data categories and descriptors mapped to GDPR-sensitive (according to GDPR Article 9) and high-risk data.} 
    \label{table:sen_schema}
    \scriptsize
    \setlength{\tabcolsep}{2pt}
    \begin{tabularx}{\linewidth}{@{} lXX @{}} 
        \hline
        Category & GDPR-sensitive descriptors & High-risk descriptors \\ 
        \hline
        Personal identifier & --- & social security number, tax ID, driver's license, passport, birth certificate, government issued ID, identity verification info \\
        Demographic info & sexual orientation, race or ethnicity, religious beliefs, political affiliation, union membership, disability status & immigration status \\
        Account info & --- & passwords, security pin, security questions and answers, MFA data, authentication token, password reset token \\
        Medical info & all & --- \\
        Biometric data & all & --- \\
        Financial info & --- & all \\
        Financial standing & --- & financial statements, credit score, credit history \\
        Insurance info & health insurance info, health insurance claims history, criminal history & --- \\
        Legal info & --- & signature, work authorization info, background screening info \\
        Precise location & --- & all \\
        Movement data & --- & all \\
        Sensory data & --- & all \\
        On-site data & --- & on-site audio/video recordings \\
        Internet usage & --- & browsing history, search history \\
        \hline
    \end{tabularx}
\end{center}

\begin{center}
    \footnotesize
    \setlength{\tabcolsep}{2pt}
    \captionof{table}{Our expert-assigned and survey-derived weights for Business Interest and User Protection. For the survey-derived weights, we report the mean and a 95\% bootstrap confidence interval for the mean.}
    \label{table:schema_weights}
    \begin{tabular}{@{} c @{\hspace{10pt}} l l c c c @{}}
    \hline
    & \multirow{2}{*}{Field} & \multirow{2}{*}{Value} & Expert & \multicolumn{2}{c}{Survey-derived} \\
    & & & weight & Mean & Mean CI \\
    \hline
    \multirow{15}{*}{\rotatebox[origin=c]{90}{Business Interest}} & \multirow{8}{*}{\texttt{purpose}} & core functionality & 4 & 4.51 & $[4.27,\ 4.60]$ \\
    & & user experience & 2 & 3.97 & $[3.76,\ 4.15]$ \\
    & & security & 1 & 4.64 & $[4.32,\ 4.66]$ \\
    & & analytics and research & 2 & 3.96 & $[3.71,\ 4.16]$ \\
    & & advertising and marketing & 3 & 3.70 & $[3.41,\ 3.94]$ \\
    & & data sale & 3 & 3.25 & $[2.88,\ 3.57]$ \\
    & & legal compliance & 1 & 4.09 & $[3.67,\ 4.28]$ \\
    & & unrestricted & 5 & 3.25 & $[2.86,\ 3.59]$ \\
    \cmidrule{2-6}
    & \multirow{7}{*}{\texttt{recipients}} & affiliate & 4 & 4.00 & $[3.75,\ 4.19]$ \\
    & & successor & 3 & 3.87 & $[3.55,\ 4.10]$ \\
    & & processor & 3 & 4.14 & $[3.81,\ 4.32]$ \\
    & & partner & 5 & 3.68 & $[3.46,\ 3.89]$ \\
    & & public authority & 2 & 3.38 & $[3.06,\ 3.66]$ \\
    & & public & 1 & 2.34 & $[2.09,\ 2.64]$ \\
    & & other & 5 & 2.99 & $[2.66,\ 3.31]$ \\
    \hline
    \multirow{23}{*}{\rotatebox[origin=c]{90}{User Protection}} & \multirow{7}{*}{\texttt{protection}} & secure authentication & 3 & 4.37 & $[4.13,\ 4.50]$ \\
    & & secure transfer & 2 & 4.40 & $[4.17,\ 4.52]$ \\
    & & secure storage & 2 & 4.52 & $[4.34,\ 4.62]$ \\
    & & access limit & 4 & 3.96 & $[3.68,\ 4.17]$ \\
    & & privacy program & 5 & 4.18 & $[3.99,\ 4.33]$ \\
    & & privacy review & 5 & 4.15 & $[3.96,\ 4.31]$ \\
    & & generic & 1 & 2.73 & $[2.42,\ 3.07]$ \\
    \cmidrule{2-6}
    & \multirow{3}{*}{\texttt{retention}} & indefinitely & 1 & 1.82 & $[1.69,\ 2.28]$ \\
    & & limited & 3 & 3.15 & $[2.90,\ 3.39]$ \\
    & & stated & 5 & 4.34 & $[4.11,\ 4.48]$ \\
    \cmidrule{2-6}
    & \multirow{5}{*}{\texttt{choice}} & no choice & 1 & 1.84 & $[1.70,\ 2.26]$ \\
    & & opt-in & 5 & 4.24 & $[3.90,\ 4.39]$ \\
    & & opt-out via link & 3 & 3.61 & $[3.35,\ 3.84]$ \\
    & & opt-out via direct contact & 2 & 3.06 & $[2.73,\ 3.37]$ \\
    & & privacy settings & 2 & 4.20 & $[3.96,\ 4.36]$ \\
    \cmidrule{2-6}
    & \multirow{8}{*}{\texttt{access}} & none & 1 & 1.73 & $[1.64,\ 2.22]$ \\
    & & view & 2 & 2.71 & $[2.42,\ 3.04]$ \\
    & & export & 5 & 3.45 & $[3.21,\ 3.69]$ \\
    & & edit & 3 & 3.55 & $[3.32,\ 3.77]$ \\
    & & restrict & 3 & 3.91 & $[3.66,\ 4.11]$ \\
    & & deactivate & 4 & 3.56 & $[3.28,\ 3.81]$ \\
    & & partial delete & 4 & 3.77 & $[3.56,\ 3.97]$ \\
    & & full delete & 5 & 4.49 & $[4.14,\ 4.56]$ \\
    \hline
    \end{tabular}
\end{center}

\newpage
\begin{center}
\centering
\scriptsize
\setlength{\tabcolsep}{2pt}
\renewcommand{\arraystretch}{1.15}
\captionof{table}{Statement pairs for which survey participants either showed no statistically significant preference ($p>0.05$) or showed a preference that contradicted the ordering implied by our Transparency weights. Votes are reported as A/B/ND, where A and B denote the two statements and ND indicates no perceived difference in transparency.}
\label{tab:disagreements}
\begin{tabular}{@{}
  l @{\hspace{5pt}}
  p{0.30\columnwidth}
  p{0.30\columnwidth}
  c
@{}}
\toprule
\textbf{} &
\multirow{2}{*}{A} & \multirow{2}{*}{B} &
Votes \\
& & & (A/B/ND) \\
\midrule

\multirow{10}{*}{\rotatebox[origin=c]{90}{Spec. (A)\ vs.\ underspec. (B)}}
& Facility is required by law to maintain the privacy of your PHI, to provide
  individuals with notice of Facility's legal duties and privacy practices with
  respect to PHI, and to abide by the terms described in this Notice.
& We may employ industry standard procedural and technological measures that are
  reasonably designed to help protect your personally identifiable information
  from loss, unauthorized access, disclosure, alteration or destruction.
& 29/16/5 \\ \addlinespace[1.5em]

\midrule

\multirow{30}{*}{\rotatebox[origin=c]{90}{Underspecified (A) vs global (B)}}
& As part of the business transfer process, we may share certain of your Personal
  Information with lenders, auditors, and third-party advisors, including
  attorneys and consultants.
& Under certain circumstances, the Company may be required to disclose Your
  Personal Data if required to do so by law or in response to valid requests by
  public authorities (e.g.\ a court or a government agency).
& 14/32/4 \\ \addlinespace[3pt]

& Your request and choices may be limited in certain cases: for example, if
  fulfilling your request would reveal information about another person, or if
  you ask to delete information which we or your administrator are permitted by
  law or have compelling legitimate interests to keep.
& You have the right to delete or request that We assist in deleting the Personal
  Data that We have collected about You.
& 25/22/3 \\ \addlinespace[3pt]

& When we use a third party service provider for a process purpose; for example,
  when paying via 42.fr to buy food, the bank data is used on the site of the
  service provider in a secure way.
& The Company will take all steps reasonably necessary to ensure that Your data
  is treated securely and in accordance with this Privacy Policy and no transfer
  of Your Personal Data will take place to an organization or a country unless
  there are adequate controls in place including the security of Your data.
& 25/25/0 \\

\midrule

\multirow{8}{*}{\rotatebox[origin=c]{90}{Clear (A) vs.\ vague (B)}}
& We may share Your information with Our affiliates, in which case we will
  require those affiliates to honor this Privacy Policy.
& Sharing information with our IAC companies enables us to provide you with
  information about a variety of products and services that might interest you.
& 22/19/9 \\ \addlinespace[2.5em]

\bottomrule
\end{tabular}
\end{center}

\newpage
\section{Example Metric Computation}\label{app:metric_computation}

The following example illustrates how our framework uses structured representations to map privacy policy text to the four metrics described in \autoref{sec:analysis}.

\begin{tcolorbox}[
    title={Example: A simple privacy policy},
    colback=white,
    colframe=gray!100,
    fonttitle=\small\bfseries,
]
\footnotesize
\textbf{Data we collect}

We collect your \emph{email address} when you create an account and automatically collect \emph{location data} from your device. 

\textbf{How we use your data} 
\begin{itemize}[leftmargin=*, noitemsep, topsep=0pt]
    \item We use your \emph{email address} to provide account-related services and send security notifications.
    \item We may also use your data for research purposes.
\end{itemize}

\textbf{Data sharing} 
\begin{itemize}[leftmargin=*, noitemsep, topsep=0pt]
    \item We may share your \emph{personal information} with service providers that help us deliver our services. 
\end{itemize}

\textbf{Data retention} 
\begin{itemize}[leftmargin=*, noitemsep, topsep=0pt]
    \item We retain your \emph{email address} for up to 30 days after your account is last active.
    \item We retain \emph{location data} only as long as necessary to support our business needs.
\end{itemize}

\textbf{Data protection} 
\begin{itemize}[leftmargin=*, noitemsep, topsep=0pt]
    \item We protect your \emph{email address} using encryption.
    \item We maintain reasonable safeguards to protect \emph{location data} collected from your device. 
\end{itemize}

\textbf{Your rights} 
\begin{itemize}[leftmargin=*, noitemsep, topsep=0pt]
    \item You may opt out of email communications by contacting us at support@example.com.
    \item You may also update your \emph{email address} through your account settings.
    \item You may manage or disable location collection through your privacy dashboard.
\end{itemize}

\end{tcolorbox}

\paragraph{Structured representation.}
From the example above, we extract two data elements ($D = \{d_1, d_2\}$) and 10 statements of data practices ($|S| = 10$), where $d_1 = \emph{email address}$ and $d_2 = \emph{location data}$. \autoref{tab:running-example} display the data practice values assigned to each data element.

\begin{table*}[t]
\caption{Structured representation of the privacy policy example after mapping text to practice values in our taxonomy, used for computing metrics.}
\label{tab:running-example}
\centering
\small
\setlength{\tabcolsep}{3pt}
\begin{tabular}{p{2cm}p{1.5cm}p{3.8cm}p{1.5cm}p{1.5cm}p{2cm}p{2cm}p{1.5cm}}
\toprule
Data element & Method & Purpose & Sharing & Retention & Protection & Choice & Access \\
\midrule
email address & direct &
core functionality; security; analytics and research &
processor &
stated &
secure storage &
opt-out via link &
edit \\
\addlinespace
location data & automatic &
analytics and research &
processor &
limited &
generic &
privacy settings &
- \\
\bottomrule
\end{tabular}
\end{table*}

\subsection{Completeness}

We compute Completeness from three sub-metrics as follows:

\[
C = \boldsymbol{\lambda}_c^\top \cdot [C_{\text{practice}}; C_{\text{desc}}; C_{\text{method}}].
\]

\paragraph{(a) Practice coverage}
For this term, we count how many of the six practices are explicitly specified for each data element. For \emph{email address}, all six practices are specified. For \emph{location data}, five out of six practices are specified, with \texttt{access} missing. Averaging across all data elements gives
\[
C_{\text{practice}}
=
\frac{1}{2}\left(\frac{6}{6} + \frac{5}{6}\right)
=
\frac{11}{12}
\approx 0.917.
\]

\paragraph{(b) Descriptor specificity}
This term captures whether the privacy policy names concrete data descriptors rather than staying at a vague category level. In this example:

\begin{itemize}
    \item \emph{email address} is a specific descriptor.
    \item \emph{location data} is treated as a category and not a specific descriptor in our schema (an example of a specific descriptor under this category is \emph{gps location}).
\end{itemize}

Thus,
\[
C_{\text{desc}} = \frac{1}{2} = 0.5.
\]

\paragraph{(c) Method specificity}
This term checks whether the privacy policy explicitly states the collection method for each data element.

\begin{itemize}
    \item \texttt{email address}: Direct collection (``when you create an account'').
    \item \texttt{location data}: Automatic collection.
\end{itemize}

Thus,
\[
C_{\text{method}} = \frac{2}{2} = 1.
\]

\subsection{Transparency}

We compute Transparency from two sub-metrics as follows:
\[
T = \boldsymbol{\lambda}_t^\top [T_{\text{scope}}; T_{\text{clarity}}].
\]

\paragraph{(a) Scope}
Each statement of a data practice is assigned a scope weight:
\[
w_{\text{scope}}(s) =
\begin{cases}
0.2, & \text{if } \text{scope}(s) = \texttt{global}, \\
0.5, & \text{if } \text{scope}(s) = \texttt{underspecified}, \\
1.0, & \text{if } \text{scope}(s) = \texttt{specified}.
\end{cases}
\]

For instance, the statement ``We use your email address to provide account-related services'' is explicitly tied to a specific data element and therefore receives a weight of $1$. In this example, all but two statements are explicitly associated with either \emph{email address} or \emph{location data}. The remaining statements, ``We may also use your data for research purposes'' and ``We may share your personal information with service providers,'' are global statements and receive a weight of $0.2$. We iterate over all statements and average their corresponding scope scores.
\[
T_{\text{scope}} = \frac{1}{10}(2 \times 0.2+8 \times 1)=0.84
\]

\paragraph{(b) Clarity}
The clarity of each statement is quantified as:
\[
w_{\text{clarity}}(s) =
\begin{cases}
0.2, & \text{if } \text{clarity}(s) = \texttt{vague}, \\
1.0, & \text{if } \text{clarity}(s) = \texttt{clear}.
\end{cases}
\]

In this example, there are two statements deemed vague: ``We maintain reasonable safeguards to \dots'' (a generic data protection statement), and ``We retain location data only as long as necessary'' (a limited but unstated retention period). Therefore, we have:
\[
T_{\text{clarity}}
= \frac{1}{10}(2 \times 0.2 + 8 \times 1) = 0.84
\]

\subsection{Business Interest}

To compute the Business Interest score, we first calculate $F(v_p)$ for each practice value $v_p$, where $F(v_p)$ is the fraction of collected data elements governed by that value. For example, for data collection purposes, the value \emph{core functionality} applies to one of the two collected data elements, namely \emph{email address}. Thus, $F(\emph{core functionality}) = 0.5$. Similarly,  $F(\emph{analytics and research}) = 1$ and $F(\emph{security}) = 0.5$, while all other purpose values receive a value of zero. We then aggregate these fractions using their corresponding weights from \autoref{table:schema_weights}:
\[
Q_{\text{purpose}} = 0.8 \times 0.5 + 0.4 \times 1 + 0.2 \times 0.5 = 0.9.
\]

We compute the sharing component in the same manner, yielding $Q_{\text{sharing}} = 0.6$, since the policy only includes sharing with data processors. The overall Business Interest score is then given by:
\[
B = \boldsymbol{\lambda}_b^\top \cdot [Q_{\text{purpose}}; Q_{\text{sharing}}].
\]

\subsection{User Protection}

The User Protection score is computed in the same manner as the Business Interest score, using the values and weights for data retention, data protection, user choices, and user access:
\[
B = \boldsymbol{\lambda}_u^\top \cdot [Q_{\text{retention}}; Q_{\text{protection}}; Q_{\text{choices}}; Q_{\text{access}}].
\]

\section{Transparency Sensitivity Analysis}
\label{app:sensitivity}

To investigate the sensitivity of the Transparency score to our choice of weights, we performed a two-dimensional grid search over the weights assigned to vague statements ($w_{\text{vague}}$) and global scopes ($w_{\text{global}}$). We varied both parameters from $0.0$ to $1.0$ to examine their joint effect on the Transparency score. As shown in Figure~\ref{fig:robust_t}, the rankings remain highly stable across a broad range of parameter combinations, with Spearman rank correlations above 0.9 relative to the baseline setting of $w_{\text{vague}} = w_{\text{global}} = 0.2$. This stability is especially evident when $w_{\text{global}} < 0.6$ and $w_{\text{vague}} < 0.8$. These results confirm that the Transparency metric is robust to small variations in these weights.

\begin{center}
    \centering
    \includegraphics[width=0.9\linewidth]{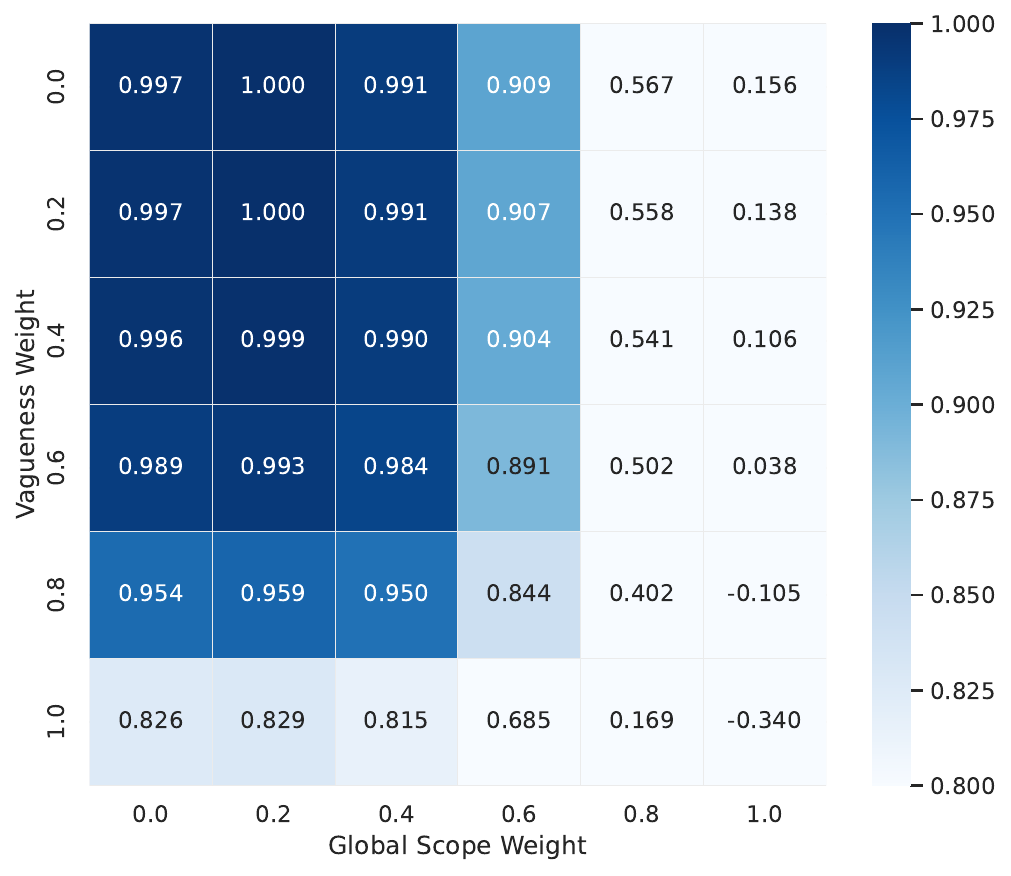}
    \captionof{figure}{Spearman correlation coefficients between the baseline Transparency scores and scores computed using alternative weights for vague ($w_{\text{vague}}$) and global ($w_{\text{global}}$) statements.}
    \label{fig:robust_t}
\end{center}

\section{Data Retention Statements from Privacy Polices A and B in \autoref{fig:pareto_frontier_communication_services}}\label{app:statemet_example}

\begin{center}
    \centering
    \begin{tcolorbox}[
      title=Context from Privacy Policy A,
      colback=white, colframe=gray!100,
      left=5pt, right=5pt, top=3pt, bottom=3pt,
      fonttitle=\footnotesize\bfseries,
    ]
        \scriptsize
        \textbf{HOW LONG DO WE KEEP YOUR PERSONAL INFORMATION?}
        \begin{itemize}[leftmargin=*, noitemsep, topsep=0pt]
            \item We retain personal information only for as long as is necessary for the purposes described in this Policy, after which it is deleted from our systems.
            \item Regarding personal information we have processed in connection with the supply of our services to clients, we may retain personal information relevant to our services for up to five years from the date of supply and in compliance with our obligations under the EU General Data Protection Regulation (or similar legislation around the world).
            \item We may then destroy such files without further notice or liability.
            \item Regarding any other personal information we have processed, we may retain relevant personal information for up to three years from the date of our last interaction with the relevant individual.
            \item We may then destroy such files without further notice or liability.
            \item If any personal information is only useful for a short period (e.g. for a specific event or marketing campaign or in relation to recruitment), we may delete it at the end of that period.
            \item If you have opted out of receiving marketing communications from us, we will need to retain certain personal information on a suppression list indefinitely so that we know not to send you further marketing communications in the future.
        \end{itemize}
    \end{tcolorbox}
\end{center}
\begin{center}
    \begin{tcolorbox}[
      title=Context from Privacy Policy B,
      colback=white, colframe=gray!100,
      left=5pt, right=5pt, top=3pt, bottom=3pt,
      fonttitle=\footnotesize\bfseries,
    ]
        \scriptsize
        \textbf{Your Choices About the Information We Collect}
        \begin{itemize}[leftmargin=*, noitemsep, topsep=0pt]
            \item To unsubscribe from any of our newsletters, follow the ``Unsubscribe'' link at the bottom of the email.
            \item Please note that certain of your personal information, such as your name or other identifying information, may remain in our database even after a deletion request in order to service your account, maintain the integrity and historical record of our database and systems, or to comply with applicable laws and regulations.
            \item You may have additional rights as set forth below depending on the jurisdiction in which you reside.
        \end{itemize}
    \end{tcolorbox}
\end{center}

\section{Example Prompts}\label{app:prompt_specifications}

\begin{tcolorbox}[
    title={Section Discovery},
    breakable,
    colback=white, colframe=gray!100,
    left=5pt, right=5pt, top=3pt, bottom=3pt,
    fonttitle=\small\bfseries,
]
\setlist{leftmargin=*, nosep}
\footnotesize
\textbf{Task}

You are a privacy policy analyst tasked with extracting sections from privacy policy texts. Your goal is to identify and extract logical sections within a given text (scraped from a webpage) that may contain a privacy policy, and to label each section based on its topic.

\textbf{Instructions}

\begin{enumerate}
    \item Read the input text:
    \begin{itemize}[leftmargin=0pt]
        \item Carefully read the text provided in the next message, which may contain a privacy policy.
        \item The text is scraped from a webpage and is formatted with zero-padded line numbers enclosed in brackets (e.g., "[023]") at the beginning of each line.
        \item The text may contain markdown formatting (e.g., headings or emphasis via bold/italic text). You can use this formatting to identify the page's structure.
    \end{itemize}
    
    \item Determine whether a privacy policy is present:
    \begin{itemize}[leftmargin=0pt]
        \item Decide whether the input text contains a privacy policy. A privacy policy must discuss at least one of the following topics:
        \begin{itemize}[leftmargin=0pt]
            \item What data is collected and how it may be collected.
            \item What are the purposes or legal bases for data collection/processing.
            \item How the collected data is handled, stored, or protected.
            \item Whether and how data is shared with or disclosed to third parties.
            \item User rights, choices, and controls with respect to their data.
        \end{itemize}
        \item If a privacy policy is present, record the first and last line numbers that mark the beginning and end of the privacy policy, then proceed to Step 3.
        \begin{itemize}[leftmargin=0pt]
            \item Exclude unrelated or boilerplate content such as site navigation (e.g., "Home", "Terms of Use"), footers (e.g., copyright notices), cookie consent banners, or contact forms.
        \end{itemize}
        \item If a privacy policy is not present, skip to Step 5.
    \end{itemize}

    \item Identify logical sections:
    \begin{itemize}[leftmargin=0pt]
        \item Within the line range identified in Step 2, determine logical sections of the privacy policy using headings or emphasis (including those indicated by markdown) and contextual cues.
        \item For each identified section, determine the following:
        \begin{itemize}[leftmargin=0pt]
            \item \texttt{"line\_numbers"}: The input line numbers for the section's first and last lines (inclusive). The first line should include the heading, if one is present.
            \item \texttt{"section\_number"}: The section's number, converting any existing numbering to dot-separated Arabic numerals (e.g., \texttt{"1"}, \texttt{"2.1"}, or \texttt{"3.2.1"}), or assigning one if none is present.
            \item \texttt{"heading"}: If the text includes a heading, capture it exactly as written, excluding markdown syntax and any section numbering. If no heading is present, create a clear and concise heading (fewer than 10 words).
        \end{itemize}
        \item Important notes:
        \begin{itemize}[leftmargin=0pt]
            \item When the privacy policy already has a clear section structure (with or without section numbers), follow that structure without splitting, merging, or creating new sections, except that you may add sections at the beginning or end to capture a table of contents or introductory/concluding text.
            \item When the privacy policy does not include its own section structure, create one. Merge contiguous ranges that cover the same topics into a single section, and do not create more than one level of nesting (e.g., allow \texttt{"1"} and \texttt{"1.1"}, but not \texttt{"1.1.1"}).
            \item Ensure that the line ranges for all sections collectively cover the entire line range identified in Step 2.
            \item Every line within the line range identified in Step 2 must be assigned to exactly one section; there must be no overlaps between sections, no gaps, and no splitting of a single line across multiple sections.
        \end{itemize}
    \end{itemize}

    \item Label sections:
    \begin{itemize}[leftmargin=0pt]
        \item For each extracted section, determine the following:
        \begin{itemize}[leftmargin=0pt]
            \item \texttt{"primary\_labels"}: The section's main topic(s), based on its heading or dominant content in the first few sentences of the section.
            \item \texttt{"secondary\_labels"}: Additional topics that are discussed in the section's body but are not its main focus.
        \end{itemize}
        \item Use the predefined enumeration in the output schema to assign the most appropriate values for entries in \texttt{"primary\_labels"} and \texttt{"secondary\_labels"}.
        \begin{itemize}[leftmargin=0pt]
            \item List labels from most to least relevant.
            \item Do not invent new topics; use only those provided in the enum.
        \end{itemize}
        \item Select a secondary label only when the section contains specific, substantive information about that topic; generic or passing references do not qualify.
    \end{itemize}

    \item Construct the output:
    \begin{itemize}[leftmargin=0pt]
        \item Your output must be a JSON-formatted string following the provided JSON schema.
        \item Each output entry must correspond to a single extracted section and include all required fields described above.
        \item If the input text does not contain a privacy policy, output an empty list (\texttt{[]}).
    \end{itemize}
\end{enumerate}

\textbf{Schema}

\dots

\textbf{Example}

\dots
\end{tcolorbox}

\begin{tcolorbox}[
    title={Extraction of collected data elements},
    breakable,
    colback=white, colframe=gray!100,
    left=5pt, right=5pt, top=3pt, bottom=3pt,
    fonttitle=\small\bfseries,
]
\footnotesize
\setlist{leftmargin=*, nosep}
\textbf{Task}

You are a privacy policy analyst tasked with extracting data collection statements from privacy policy texts. Your goal is to extract collected data elements, map them to predefined categories and normalized descriptors, and enrich them with additional contextual attributes.

\textbf{Instructions}

\begin{enumerate}[leftmargin=1.5em, labelsep=0.5em, label=\textbf{(\arabic*)}]
    \item Read the input:
    \begin{itemize}[leftmargin=0pt]
        \item Carefully and thoroughly read the text provided in the next message, which contains sections of a privacy policy related to data collection.
        \item The text is formatted with zero-padded line numbers enclosed in brackets (e.g., "[023]") at the beginning of each line.
    \end{itemize}

    \item Extract collected data elements:
    \begin{itemize}[leftmargin=0pt]
        \item Identify every phrase that explicitly describes a data element being collected, inferred, or received by the first party.
        \begin{itemize}[leftmargin=0pt]
            \item The first party is the organization that the privacy policy belongs to.
            \item Ignore statements where the first party is not the recipient of the data (e.g., statements describing the first party sharing data with other entities).
        \end{itemize}
        \item Extract the exact phrase from the text (\texttt{"phrase"}) that describes the data element.
        \begin{itemize}[leftmargin=0pt]
            \item You may capture discontinuous words if needed, but do not introduce new wording or paraphrase.
        \end{itemize}
        \item Record the line number from the input (\texttt{"line\_index"}) where the phrase appears.
        \begin{itemize}[leftmargin=0pt]
            \item Keep each phrase within a single line whenever possible.
            \item Use multiple lines only when the phrase's meaning is clearly split across them; in that case, record the starting line number.
        \end{itemize}
        \item Important notes:
        \begin{itemize}[leftmargin=0pt]
            \item Capture every mention of each data element, even if it is repeated multiple times in the text.
            \item Capture only the minimal text needed to clearly identify the data element, excluding any surrounding or irrelevant text.
            \item Each captured phrase must refer to exactly one data element. If a statement mentions multiple elements, split it into separate entries.
            \item Focus strictly on what data is being collected, not on how it is collected, used, or shared.
            \item Ignore hypothetical or negated statements about data collection (e.g., "we do not collect …").
        \end{itemize}
    \end{itemize}

    \item Categorize and normalize data elements:
    \begin{itemize}[leftmargin=0pt]
        \item For each extracted phrase, determine the following:
        \begin{itemize}[leftmargin=0pt]
            \item \texttt{"category"}: A standard category for the data element.
            \begin{itemize}[leftmargin=0pt]
                \item Use the predefined enumeration in the output schema to assign the most appropriate value.
                \item If a data element matches multiple categories, select the most relevant one.
                \item If a data element does not match any of the specific categories in the enum, set \texttt{"category"} to \texttt{"Other"}.
            \end{itemize}
            \item \texttt{"descriptor"}: A normalized descriptor for the data element (to map different wordings to a common terminology).
            \begin{itemize}[leftmargin=0pt]
                \item Use the predefined enumeration for the selected data category in the output schema to assign the most appropriate value.
                \item If a data element matches multiple descriptors, select the most specific/relevant one.
                \item If the phrase refers only to a broad data category (e.g., "contact information") and does not identify any concrete data element within that category, set \texttt{"descriptor"} to \texttt{"category-level"}.
                \item If the phrase identifies a concrete data element (i.e., it is more detailed than a broad category) but does not match any of the specific descriptors in the enum, set \texttt{"descriptor"} to \texttt{"other"}.
            \end{itemize}
            \item \texttt{"descriptor\_other"}: When \texttt{"descriptor"} is \texttt{"other"}, use this field to provide your own normalized descriptor, following the naming conventions used in the \texttt{"descriptor"} enum.
        \end{itemize}
    \end{itemize}

    \item Assign contextual attributes:
    \begin{itemize}[leftmargin=0pt]
        \item For each extracted phrase, use context from the surrounding text to determine:
        \begin{itemize}[leftmargin=0pt]
            \item \texttt{"subjects"}: The individuals or entities from whom the data element is being collected.
            \item \texttt{"collectors"}: The entities that initially collect the data element (i.e., the original collectors).
            \item \texttt{"methods"}: The methods by which the data element is initially collected or obtained from the perspective of \texttt{"collectors"}.
        \end{itemize}
        \item Use the predefined enumerations in the output schema to select all applicable values for \texttt{"subjects"}, \texttt{"collectors"}, and \texttt{"methods"}.
        \begin{itemize}[leftmargin=0pt]
            \item Carefully read the description of each enum value to determine applicable matches.
            \item Select only values supported by the current phrase and its immediate context; ignore values mentioned elsewhere in the text (those will be captured with their own phrases).
        \end{itemize}
    \end{itemize}

    \item Construct the output:
    \begin{itemize}[leftmargin=0pt]
        \item Your output must be a JSON-formatted string following the provided JSON schema.
        \item Each output entry must correspond to a single extracted phrase and include all required fields described above.
        \item If the input text does not mention any collected data elements, output an empty list (\texttt{[]}).
    \end{itemize}
\end{enumerate}

\textbf{Schema}

\dots

\textbf{Example}

\dots
\end{tcolorbox}

\begin{tcolorbox}[
    title={Extraction of data collection purposes},
    breakable,
    colback=white, colframe=gray!100,
    left=5pt, right=5pt, top=3pt, bottom=3pt,
    fonttitle=\small\bfseries,
]
\footnotesize
\setlist{leftmargin=*, nosep}
\textbf{Task}

You are a privacy policy analyst tasked with extracting statements describing the purposes for which data is collected or used from privacy policy texts. Your goal is to identify and extract structured attributes from statements describing the purposes or legal bases for data collection, and to link each statement to previously extracted data elements and subjects.

\textbf{Instructions}
\begin{enumerate}[leftmargin=*]
    \item \textbf{Read the input}: You will receive two parts as input:
    \begin{itemize}[leftmargin=0pt]
        \item Privacy policy text:
        \begin{itemize}[leftmargin=0pt]
            \item Carefully and thoroughly read the text provided, which contains sections of a privacy policy related to data collection purposes.
            \item The text is formatted with zero-padded line numbers enclosed in brackets (e.g., ``[023]'') at the beginning of each line.
        \end{itemize}
        \item Collected data elements: Review the provided JSON-formatted string containing data elements described as being collected within the same privacy policy, along with the subjects they are collected from.
        \begin{itemize}[leftmargin=0pt]
            \item \texttt{"contexts"} (list of strings): The specific sentences from the privacy policy from which the collected data elements are extracted.
            \item \texttt{"data\_elements"} (list of dictionaries): The collected data elements (and their corresponding subjects) extracted from \texttt{"contexts"}. Each item includes:
            \begin{itemize}[leftmargin=0pt]
                \item \texttt{"category"} (string): The standard data category assigned to the data element.
                \item \texttt{"descriptor"} (string): A granular, normalized data descriptor (within the data category) assigned to the data element.
                \item \texttt{"phrases"} (list of strings): The exact phrases extracted from \texttt{"contexts"} that describe the data element.
                \item \texttt{"subjects"} (list of strings): The individuals or entities from whom the data element is being collected.
            \end{itemize}
        \end{itemize}
    \end{itemize}

    \item \textbf{Extract data collection purposes}:
    \begin{itemize}[leftmargin=0pt]
        \item Identify every statement that explicitly describes a purpose or legal basis for which data is collected or used.
        \item Extract the exact phrase from the text (\texttt{"phrase"}) that describes the purpose.
        \begin{itemize}[leftmargin=0pt]
            \item You may capture discontinuous words if needed, but do not introduce new wording or paraphrase.
        \end{itemize}
        \item Record the line number from the input (\texttt{"line\_index"}) where the phrase appears.
        \begin{itemize}[leftmargin=0pt]
            \item Keep each phrase within a single line whenever possible.
            \item Use multiple lines only when the phrase's meaning is clearly split across them; in that case, record the starting line number.
        \end{itemize}
        \item Important notes:
        \begin{itemize}[leftmargin=0pt]
            \item Capture every mention of each purpose, even if it is repeated multiple times in the text.
            \item Capture only the minimal text needed to clearly identify the purpose, excluding any surrounding or irrelevant text.
            \item Each captured phrase must refer to exactly one purpose. If a statement mentions multiple purposes, split it into separate entries.
            \item Include statements that explicitly rule out or exclude a purpose (e.g., ``we do not use your data for \ldots'').
            \item Ignore purely hypothetical statements that do not describe purposes that are actually used or implemented.
        \end{itemize}
    \end{itemize}

    \item \textbf{Link purposes to extracted data elements and subjects:}
    \begin{itemize}[leftmargin=0pt]
        \item For each extracted purpose, use the input text and the provided data elements to determine the following.
        \begin{itemize}[leftmargin=0pt]
            \item \texttt{"related\_data"}: All related data elements from the input list that the purpose applies to. Use the fields below to best match the purpose to its associated set of data.
            \begin{itemize}[leftmargin=0pt]
                \item \texttt{"is\_global"}: Set to \texttt{True} if the purpose applies globally across all collected data; set to \texttt{False} otherwise.
                \item \texttt{"is\_underspecified"}: Set to \texttt{True} if the purpose applies to a subset of collected data, but the text does not clearly identify the full set of related data elements; set to \texttt{False} otherwise.
                \item \texttt{"is\_non\_identifiable"}: Set to \texttt{True} if the purpose applies only to data described as de-identified, anonymized, or aggregated; set to \texttt{False} otherwise.
                \item \texttt{"categories"}:
                \begin{itemize}[leftmargin=0pt]
                    \item If \texttt{"is\_global"} is \texttt{False}, include all data categories from the input list that the purpose applies to as a whole (i.e., the purpose covers all descriptors under those categories).
                    \item If \texttt{"is\_global"} is \texttt{True}, or if no entire category is covered by the purpose, set this field to \texttt{None}.
                \end{itemize}
                \item \texttt{"descriptors"}:
                \begin{itemize}[leftmargin=0pt]
                    \item If \texttt{"is\_global"} is \texttt{False}, include all data descriptors from the input list that the purpose applies to.
                    \item If \texttt{"is\_global"} is \texttt{True}, or if no specific descriptor is covered by the purpose, set this field to \texttt{None}.
                    \item You do not need to include descriptors whose categories are already in \texttt{"categories"}; only list descriptors that are explicitly covered in addition to entire categories.
                \end{itemize}
            \end{itemize}
            \item \texttt{"related\_subjects"}: All related subjects from the input list that the purpose applies to, or \texttt{"global"} if the purpose applies globally regardless of the subject.
        \end{itemize}
        \item Important notes:
        \begin{itemize}[leftmargin=0pt]
            \item Cross-reference the text with the provided \texttt{"contexts"} to identify the relevant data elements. For example, if the text mentions ``information you provide when you create an account'', use \texttt{"contexts"} to determine exactly what that includes (if specified).
            \item You can use \texttt{"categories"} together with \texttt{"descriptors"}. For example, a purpose might apply to all data under one category (include that category in \texttt{"categories"}), as well as specific descriptors under another category (include those descriptors in \texttt{"descriptors"}).
            \item Do not infer categories, descriptors, or subjects. Include only what is explicitly named in the text or is clearly supported by \texttt{"contexts"}.
            \item Do not invent new categories, descriptors, or subjects; use only those provided in the input list.
        \end{itemize}
    \end{itemize}

    \item \textbf{Assign structured attributes}:
    \begin{itemize}[leftmargin=0pt]
        \item For each extracted phrase, determine the following:
        \begin{itemize}[leftmargin=0pt]
            \item \texttt{"category"}: The associated category of the purpose. Use the predefined enumeration in the output schema to assign the most appropriate value.
            \item \texttt{"negated"}: Set to \texttt{True} if the phrase describes the absence or exclusion of a purpose (either globally or for a specific subset of collected data); set to \texttt{False} otherwise.
        \end{itemize}
    \end{itemize}

    \item \textbf{Construct the output}:
    \begin{itemize}[leftmargin=0pt]
        \item Your output must be a JSON-formatted string following the provided JSON schema.
        \item Each output entry must correspond to a single extracted purpose and include all required fields described above.
        \item If the input text does not mention any data collection purposes, output an empty list (\texttt{[]}).
    \end{itemize}
\end{enumerate}

\textbf{Schema}

\dots

\textbf{Example}

\dots

\end{tcolorbox}

\end{document}